\documentclass[aps,prb,twocolumn,showpacs]{revtex4-1}
\usepackage[utf8]{inputenc}
\usepackage[american,british]{babel}
\usepackage[T1]{fontenc}
\usepackage[pdftex]{graphicx} 
\usepackage{graphicx, xcolor}
\usepackage{dcolumn}
\usepackage{bm}
\usepackage{amsmath,amsthm,amssymb}
\usepackage{color}
\usepackage{verbatim}

\definecolor{darkGreen}{RGB}{0,110,0}
\definecolor{darkBlue}{RGB}{0,0,130}
\usepackage[colorlinks,citecolor=blue,linkcolor=darkBlue,urlcolor=blue,hyperindex]{hyperref}

\usepackage{mathrsfs}
\usepackage{mathtools}
\usepackage{amsfonts}
\usepackage{amsthm}
\usepackage{amssymb}

\DeclareMathOperator{\tr}{Tr}

\newcommand{\bra}[1]{\langle #1|}
\newcommand{\ket}[1]{|#1\rangle}

\newcommand{\be}{\begin{equation}}
\newcommand{\ee}{\end{equation}}

\renewcommand{\)}{\right)}

\usepackage{bm}

\begin{document}

\title{Entanglement Hamiltonian of quantum critical chains and conformal field theories}



\author{T. Mendes-Santos$^{1}$}
\author{G. Giudici$^{1,2,3}$}
\author{M. Dalmonte$^{1,2}$}
\author{M. A. Rajabpour$^{4}$ }
\affiliation{$^{1}$The Abdus Salam International Centre for Theoretical Physics, strada Costiera 11, 34151 Trieste, Italy}
\affiliation{$^{2}$SISSA,via Bonomea, 265, 34136 Trieste, Italy}
\affiliation{$^{3}$INFN, sezione di Trieste,  34136 Trieste, Italy}
\affiliation{$^{4}$Instituto de Fisica, Universidade Federal Fluminense,
Av. Gal. Milton Tavares de Souza s/n, Gragoat\'a, 24210-346, Niter\'oi, RJ, Brazil}

\begin{abstract}

We consider a lattice version of the Bisognano-Wichmann (BW) modular Hamiltonian as an ansatz for the bipartite entanglement Hamiltonian of the quantum critical chains.
Using numerically unbiased methods, we check the accuracy of the BW-ansatz by both comparing the BW R\'enyi entropy to the exact results, and by investigating the size 
scaling of the norm distance between the exact reduced density matrix and the BW one.
Our study encompasses a variety of models, scanning different universality classes, including integrable models such as 
the transverse field Ising, three states Potts and XXZ chains and non-integrable bilinear-biquadratic model. 
We show that the R\'enyi entropies obtained via the BW ansatz properly describe the scaling properties predicted by conformal field theory.
Remarkably, the BW R\'enyi entropies faithfully capture also the corrections to the conformal field theory scaling associated 
to the energy density operator. In addition, we show that the norm distance between the discretized BW density matrix and the 
exact one asymptotically goes to zero with the system size: this indicates that the BW-ansatz can be also employed to predict properties of the eigenvectors of 
the reduced density matrices, and is thus potentially applicable to other entanglement-related quantities such as negativity.

\end{abstract}

\maketitle

\section{Introduction}
Over the last years, bipartite R\'enyi entropies (REs) have become a paradigmatic quantity in the characterization of quantum many-body lattice models \cite{Amico2008,Eisert2010,Laflorencie2016}.
In quantum critical chains, for instance, the scaling of the bipartite R\'enyi entropy of the ground state is associated with the underlying conformal field theory (CFT)~\cite{HLW1994,Vidal2003,Calabrese2004} describing its low energy properties. When the subsystem consists of a single, simply connected interval, the leading and the subleading scaling of the
R\'enyi entropy with the subsystem size give access to  the corresponding central charge~\cite{HLW1994,Vidal2003,Calabrese2004} and the partial 
operator content \cite{Cardy2010,Calabrese2010,Dalmonte2011,Alcaraz2011,Alcaraz2012} of the 
theory, respectively.
In more than one-dimension, REs also determine universal properties of the system, e.g., number of Goldstone modes~\cite{Metlitski2011} in spontaneously-symmetry-broken phases, 
and serve as a diagnostic tool to characterize topologically ordered phases~\cite{2006PhRvL..96k0405L,Kitaev_2006,Flammia:2009aa}. 

From a quantum information perspective, REs characterize the entanglement between a subsystem $A$ of a  
pure state (here we focus on the ground states and simply connected subsystems), $\ket{\psi_{AB}}$, and its complement $B$.
In the following we consider the $\alpha-$R\'enyi entropies, defined as:
\begin{align}
 S_{\alpha} = \frac{1}{1-\alpha} \ln \tr{ \rho_{A}^{\alpha}},
 \label{defS}
\end{align}
where $\rho_{A} = \tr_{B}\ket{\psi_{AB}}\bra{\psi_{AB}}$; with $\tr_{B}$ being the trace over the complement of $A$. 
For the limiting case $\alpha=1$, one obtains the von Neumann entropy, that is, the bipartite entanglement measure for 
pure states. REs with $\alpha>1$ provide strict lower bounds on the entanglement between A and B. 
Below, we are interested in cases where it is possible to uniquely determine the logarithm of the reduced density matrix:
\begin{align}
\rho_{A} =e^{-\tilde{H}_A},
\label{rdm}
\end{align}
where the operator $\tilde{H}_A$ is known as the  modular Hamiltonian in the quantum field 
theory community \cite{bisognano1975duality,bisognano1976duality,hislop1982,BGL1993,Casini:2011aa,Witten2018} and the entanglement Hamiltonian (EH)
in the condensed matter physics one~\cite{Haldane2008,Regnault:2015aa,Laflorencie2016}. For pure states, the spectrum of the EH uniquely characterizes the entanglement properties of the system.
The knowledge of a functional form of the EH is of tremendous importance.
From the experimental side, it allows to measure the entanglement
properties of a given state via direct engineering of the
EH. In particular, the EH is very useful in cases where direct access to the
wave function is not scalable, such as in experiments
requiring full state tomography~\cite{Dalmonte2017}.
From the theoretical side, the EH allows to characterize $\rho_{A}$ as a thermal state,
which opens up the possibility to investigate the R\'enyi entropy
using conventional statistical mechanics techniques \cite{Assad2018,Mendes2019} and to formulate search algorithms for parent Hamiltonians~\cite{Turkeshi_2019}.

Recently, it has been proposed~\cite{Dalmonte2017,Giuliano2018} that the EH corresponding to the ground state of the lattice models whose low-energy physics is captured by a
Lorentz invariant field theory can be approximated by a lattice adaptation of the Bisognano-Wichmann (BW) theorem~\cite{bisognano1975duality,bisognano1976duality,hislop1982}.
As discussed in Refs.~\onlinecite{Peschel2004,Kim_2016,1806.08060,Assad2018,Dalmonte2017,Kosior:2018aa,Giuliano2018,Turkeshi_2019,Peschel2018},
the corresponding reduced density matrix, $\rho_{BW}$, although not generically exact, accurately reproduces not just the low-lying 
entanglement spectrum, but also properties directly related to its eigenvectors, 
such as correlation functions and order parameters~\cite{Giuliano2018}. Furthermore, the von Neumann entropy obtained from $\rho_{BW}$, i.e. $S_{1}^{BW}$,
accurately describes universal properties, such as, the central charge of one-dimensional critical models \cite{Mendes2019,Peschel2018}.

In this work, we provide a systematic investigation of the accuracy of the BW R\'enyi entropy in the context of quantum critical chains. 
Our investigation focus on lattice models belonging to different universality classes, including the XXZ model, the transverse field Ising chain, the three-state Potts and the bilinear-biquadratic model.
The low-energy degrees of freedom of these models are described by a CFT that is characterized by a central charge $c$. 
The main issues we address here are the following: we investigate whether the R\'enyi entropy obtained from $\rho_{BW}$ is able to describe
(i) universal, leading contributions related to the central charge, (ii) non-universal terms of $S_{\alpha}$, e.g., additive constants, 
and (iii) lattice-finite-size contributions of $S_{\alpha}$ that are related to universal properties \cite{Calabrese2010,Dalmonte2011,Alcaraz2012}.
While point (iii) shall not affect the asymptotic behavior of the BW R\'enyi entropy, a negative answer for point (ii) can imply that $S_{\alpha}^{BW}$ is not exact even in the thermodynamic limit. 
We remark that since the BW-EH is based on a quantum field theory result, one expects that, while universal properties should be well captured, 
non-universal ones and contributions due to lattice-finite-size effects are not necessarily captured by $\rho_{BW}$.
Finally, (iv) we query about the accuracy of the BW reduced density matrix itself (i.e., both eigenvalue and eigenvector properties) by investigating its norm distance with the exact reduced density matrix.

We carry out a direct comparison between the exact and the BW results by combining different numerical techniques, including exact diagonalization, density-matrix-renormalization-group (DMRG) and 
quantum Monte Carlo (QMC). The main conclusions drawn from our numerical analysis are the same for all models studied, and are summarized as follows: 
(i) the R\'enyi entropy obtained from the $\rho_{BW}$, $S_{\alpha}^{BW}$, converges to the exact one in the thermodynamic limit;
(ii) $S_{\alpha}^{BW}$ properly describes not just the logarithmically-divergent CFT term, but
also corrections to the CFT scaling related to universal quantities (e.g., operator content of the theory);
and (iii) we observe that the Schatten distance between $\rho_{BW}$ and the exact reduced density matrix asymptotically goes to zero. Overall, these results point to the fact that the predictive power of the BW-EH goes well beyond what is naively expected for typical field theory expectations, thus considerably extending its applicability window.

The structure of the paper is as follows: in Sec.~\ref{method}, we present the theoretical background necessary to discuss our results, which includes a discussion about the lattice version of the 
BW-EH, and a review of the general behavior of the R\'enyi entropy in quantum critical chains.
In Sec.~\ref{resultsI} we present our comparison between the  BW and the exact results of the R\'enyi entropy, while in Sec.~\ref{resultsII} 
we discuss the behavior of the norm distance between the BW and the exact reduced density matrices. Finally, in Sec.~\ref{conclusion} we present our
main conclusions and connect them with other related works.

\section{Background}
\label{method}

In this section we provide some background material on the BW theorem, and its adaptation to lattice models.
Further we discuss how to compute R\'enyi entropies from the thermodynamic properties of the EH, and the general behavior of the R\'enyi entropies in the quantum critical chains.
Finally, we describe the quantities analyzed in the subsequent sections.

\subsection{Bisognano-Wichmann entanglement Hamiltonian on the lattice}
\label{BWlattice}

\begin{figure}[]
{\centering\resizebox*{8.5cm}{!}{\includegraphics*{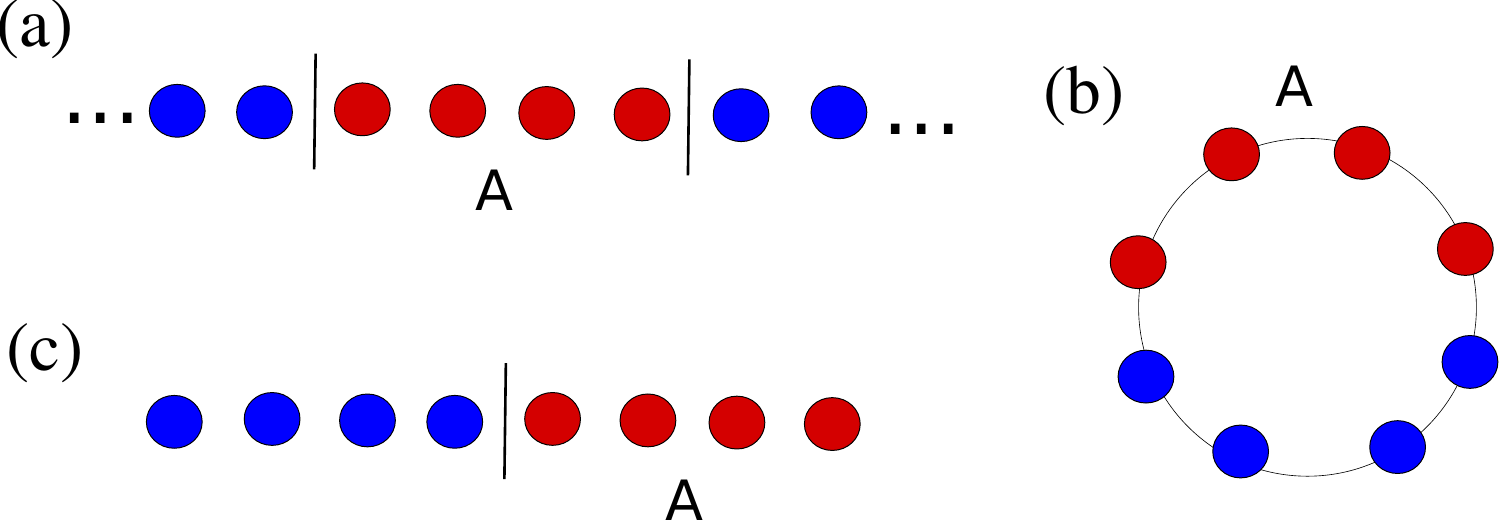}}}
\caption{\textit{Partitions of the one-dimensional systems that we consider}: 
(a) partition of length L embedded in an infinite system (infinite PBC); 
(b) half-partition of a ring (finite PBC), (c) half-partition of an open system (finite OBC).
The BW couplings of these systems are given by the CFT generalization of the BW theorem (see text).}
\label{fig:cartoon} 
\end{figure}

In the context of relativistic quantum field theories, 
there are special cases in which the entanglement (modular) Hamiltonian (EH) can be expressed as an integral over the local Hamiltonian-density, $H(\vec{x})$,
with a suitable weight factor, i.e.,
\begin{align}
\tilde{H}_A = 2 \pi \int_{\vec{x} \in A} d\vec{x}\left(\lambda(\vec{x}) H(\vec{x}) \right) + c^{\prime},
\label{BWtheorem}
\end{align}
where $c^{\prime}$ is a constant to guarantee unit trace of the density matrix, and the speed of light has been set to unity.
For example, when $A$ corresponds to a half-bipartition of an infinite system in the vacuum (ground state), 
[i.e. $\vec{x} \equiv (x_1,x_2,...,x_d) \in R^d$ and $A = {\vec{x} |x_1 > 0}$],
the Bisognano-Wichmann (BW) theorem states that $\lambda(\vec{x})~=~x_1$ \cite{bisognano1975duality,bisognano1976duality}.
For conformal field theories (CFTs), the BW theorem can be generalized to different geometries of the partition~\cite{hislop1982,Casini:2011aa,Wong2013,Cardy2016,NR2016} (see Fig.~\ref{fig:cartoon}).

The EH obtained from the discretization of the BW theorem (and its generalizations for CFT systems) has been recently explored
in different lattice models \cite{Dalmonte2017,Giuliano2018,Peschel2018,Turkeshi_2019}.
For a generic  one-dimensional
Hamiltonian composed simply of one- and two-body operators, i.e.,
\begin{align}
 H =   \Gamma \sum_{n=1}^{L_{T}}  \hat{h}_{n,n+1} + \Theta \sum_{n=1}^{L_{T}} \hat{l}_{n},
 \label{EHBW}
\end{align}
where $L_T$ is the size of the system, $\Gamma$ is a coupling (e.g., exchange term) and $\Theta$ is an on-site term (e.g., magnetic field),
we define the BW-EH as
\begin{align}
 \tilde{H}_{BW} = \beta_{BW} \left[ \Gamma \sum_{n=1}^{L-1} \lambda(n)  \hat{h}_{n,n+1} +   \Theta \sum_{n=1}^{L} \lambda(n-1/2) \hat{l}_{n} \right],
 \label{EHBW2}
\end{align}
where $L$ is the size of the subsystem, and the 
function $\lambda(n)$ defines the geometry of the  partition. 
For instance, for half-infinite partition, one has  
\begin{align}
 \lambda(n) = n,
 \label{infiniteOBC}
\end{align}
which corresponds to the lattice version of the BW theorem Eq.\,\eqref{EHBW}.

In this work, we focus on partitions with the geometries shown in the Fig. \ref{fig:cartoon}.
The corresponding BW-EH  can be obtained for  CFT systems with the appropriate choice of the couplings $\lambda(n)$.
For a subsystem that is embedded in an infinite system, one has\cite{Casini:2011aa}
\begin{align}
 \lambda(n) = \frac{n(L-n)}{L}.
 \label{infinitePBC}
\end{align}
For finite systems with half-bipartition, i.e. $L = L_{T}/2$, in the case of open and periodic boundary conditions one has\cite{Cardy2016} 
 \begin{align}
 \lambda(n) = \frac{2L}{\pi} \sin\left(\frac{\pi n}{2L} \right),
 \label{finiteOBC}
\end{align}
and 
\begin{align}
 \lambda(n) = \frac{L}{\pi} \sin\left(\frac{\pi n}{L} \right),
 \label{finitePBC}
\end{align}
respectively. We call these partitions finite OBC and finite PBC, respectively.

The other important parameter of the BW-EH is its \textit{overall} energy scale, that is conveniently defined as the BW inverse temperature
\begin{align}
 \beta_{BW} = \frac{2\pi}{v},
 \label{beta}
\end{align}
where $v$ is the sound velocity of the corresponding quantum field theory. The notation as an inverse temperature provides a simple analogy between the reduced density matrix and a partition function.

Finally, we define the BW reduced density matrix as
\begin{align}
 \rho_{BW} = \frac{e^{-\beta_{BW} H_{BW}}}{Z_{BW}},
 \label{reducedBW}
\end{align}
where the constant $Z_{BW} = \tr{e^{-\beta_{BW} H_{BW}}}$ guarantees the proper normalization of the BW reduced density matrix, $\tr(\rho_{BW}) = 1$. 
For later convenience we define $H_{BW} = \tilde{H}_{BW}/\beta_{BW}$.

\subsection{R\'enyi entropy via the Entanglement Hamiltonian }
\label{renyiBW}

Let us now discuss how to obtain the R\'enyi entropy with
the aid of the BW-EH.
If we substitute the definition of the BW reduced matrix, Eq.\eqref{reducedBW}, in
Eq. \eqref{defS}, we obtain
\begin{align}
 S_{\alpha}^{BW} = \frac{\alpha \beta_{BW}}{1-\alpha} \left[ F(\beta_{BW}) - F(\alpha \beta_{BW})  \right],
 \label{BWrenyi}
\end{align}
where $F(\beta_{BW}) = -\frac{1}{\beta_{BW}} \ln{Z_{BW}}$ is equal to the free energy of the BW-EH
at the inverse entanglement temperature, $\beta_{BW}$. 
In the limit $\alpha \to 1$, $S_{\alpha}$ reduces to the the von Neumann entropy, and 
$S_{1}^{BW}$ (von Neumann BW entropy) is equal to
\begin{align}
 S_{1}^{BW} = \beta_{BW} \left< H_{BW} \right> + \ln Z_{BW},
 \label{BWvonneumann}
\end{align}
which is nothing else but the definition of the thermodynamic entropy of the BW-EH system at $\beta_{EH}$.

Both the BW R\'enyi entropy with $\alpha > 1$ and the von Neumann entropy 
can be obtained by computing the thermodynamic properties of the BW-EH.
For instance, for a  quadratic fermionic EH, we can write 
\begin{align}
 H_{q} = \sum_{k} \epsilon(k) \hat{c}_{k}^{\dagger} \hat{c}_k,
 \label{eq:quadratic}
\end{align}
where $\epsilon(k)$ is the single-particle spectrum of $H_q$.
One then can simply employ the conventional definition of the free-energy 
for non-interacting fermions
\begin{align}
F(\beta) =  -\frac{1}{\beta} \sum_{k} \ln\left[e^{-\beta\epsilon(k)} + 1\right],
\label{eq:freeQ}
\end{align}
and use Eq. \eqref{BWrenyi} to compute $S_{\alpha}^{BW}$. 
We use this expression to compute $S_{\alpha}^{BW}$ for both the XX and the transverse field Ising models, see Eqs. \eqref{modelXX} and \eqref{modelIsing}, respectively.
As discussed in the Appendix \ref{appendixA}, the BW-EH of these models can be cast in the same form as of Eq. \eqref{eq:quadratic}. 

For models whose Hamiltonians cannot be cast in a quadratic form,
one can use quantum Monte Carlo methods to compute $S_{\alpha}^{BW}$.
It is important to mention that the BW-EH of sign-problem-free models, as the ones considered here, is also  sign-free.
For sign-problem models, one can use tensor network methods to compute $S_{\alpha}^{BW}$.
We consider the quantum version of the Wang-Landau method \cite{Wang2001}
performed in the stochastic series expansion (SSE) framework \cite{Wessel2003}, \cite{Wessel2007}.
This method allows a direct calculation of the free-energy and the entropy of the BW-EH at $\beta = \beta_{EH}$.
Here we use both local and SSE directed-loop updates to simulate the XXZ model \cite{sandvik1991}, \cite{sandvik2002}.
Using WL-QMC we can compute $S_{\alpha}^{BW}$ for system sizes comparable with the ones 
achieved with DMRG, $L \sim 10^{2}$. The method can also be straightforwardly applied to dimension $d>1$~\cite{Mendes2019}. 
For interacting systems, we also employ exact diagonalization (ED) methods to compute $S_{\alpha}^{BW}$.

We compare  $S_{\alpha}^{BW}$ with the R\'enyi entropy obtained with 
the exact reduced density matrix, $\rho_A$.
From now on we call the \textit{exact} R\'enyi entropy $S_{\alpha}$.
The general behavior of $S_{\alpha}$ for quantum critical chains is described in the next section.
All the calculations of $\rho_A$ are performed with exact numerical methods. 
For the non-interacting systems i.e., XX and transverse field Ising models, we obtain $\rho_A$ with 
the aid of the correlation matrix \cite{Peschel2004},
and for interacting systems we use both ED and DMRG methods.

\subsection{R\'enyi entropy in the quantum critical chains}
\label{renyiCFT}

The R\'enyi entropy of the ground state of one-dimensional models whose low-energy physics is captured by a gapless relativistic field theory has been 
extensively studied by both analytical and numerical methods, see, e.g., Ref.~\onlinecite{Amico2008,CCD2009,Eisert2010,Laflorencie2016} for reviews.
There are numerous analytical and numerical results indicating that the leading asymptotic behavior of  $S_{\alpha}$ for $\alpha \to 1^+$
coincides with the entropy of the vacuum in the CFTs~\cite{CCD2009}, i.e. for $L/a \gg 1$, ($a$ is the lattice constant), $S_{\alpha} = S_{\alpha}^{CFT}$.
For example, when the  subsystem is a single interval of length $L$ embedded in an infinite system [see Fig.\ref{fig:cartoon} (a)], one has\cite{HLW1994} 
\begin{align}
 S_{\alpha}^{CFT} (L)= \frac{c}{6} \left( 1 + \frac{1}{\alpha} \right) \ln \frac{L}{a}  + c_{\alpha}^{\prime},
 \label{Scft}
\end{align}
where $c$ is the central charge of the corresponding CFT and  $c_{\alpha}^{\prime}$ is a non-universal constant.
The CFT formula is also generalized for a finite system with the length $L_T$ where we have\cite{HLW1994,Calabrese2004} 
\begin{align}
 S_{\alpha}^{CFT} (L,L_T)= \frac{c}{6\eta} \left( 1 + \frac{1}{\alpha} \right) \ln \left[\frac{\eta L_T}{\pi a} \sin\frac{\pi L}{L_T} \right]  + c_{\alpha}^{\prime\prime},
 \label{ScftFinite}
\end{align}
where $\eta = 1,2$ for PBC/OBC, and $c^{\prime\prime}_{\alpha}$ is a non-universal constant.

Away from the asymptotic limit, i.e. $L\approx a$, (from now on we use $a=1$) it is known that $S_{\alpha}$ is also dominated by  corrections to the CFT expressions, i.e.,  
\begin{equation}
 C_{\alpha}(L) = S_{\alpha}(L) - S_{\alpha}^{CFT}(L) \neq 0.
\end{equation}
As first noticed in the Ref.~\onlinecite{Laflorencie2006} for systems with OBC, the CFT formula cannot explain the oscillations  observed in the von Neumann entropy.
In the Ref.~\onlinecite{Calabrese2010}, it was observed that parity oscillations in the $S_{\alpha}$ of the XXZ model can also occur in a system with PBC for $\alpha > 1$. There 
it was proposed that the asymptotic leading term of $C_{\alpha}(L)$  is given by
\begin{align}
 d_{\alpha}(L) = f_{\alpha} \cos \left(2 k_{F} L \right) \left|  2 L \sin(k_F)    \right|^{-p_{\alpha}},
 \label{d}
\end{align}
where $f_{\alpha}$ is a nonuniversal constant, $p_{\alpha}$ is a universal critical exponent equal to $p_{\alpha} = 2K/\alpha$ and
$K$ is the  Luttinger liquid parameter.
For the U($1$) XXZ model, $k_{F}^{XXZ} = \pi/2$, and $d_{\alpha}(L)$  oscillates with $L$.
In this case, the presence of $d_{\alpha}(L)$ is confirmed by both exact numerical calculations based on DMRG \cite{Laflorencie2006}, and
the exact analytical solution of the R\'enyi entropy of the XX model\cite{Calabrese2010_2}.
The oscillatory behavior of these corrections is attributed to the 
the tendency to antiferromagnetic order in the ground state of the XXZ model.
Oppositely, for the transverse field Ising model, as $k_{F}^{Ising} = 2k_{F}^{XXZ}$,  the leading term of $C_{\alpha}(L)$ is given by a nonoscillatory $d_{\alpha}(L)$\cite{Calabrese2010}.

The fact that the asymptotic leading term of $C_{\alpha}(L)$ is equivalent to $d_{\alpha}(L)$ has
been confirmed with DMRG calculations for models belonging to different universality classes 
in finite systems with both PBC and OBC \cite{Laflorencie2006,Dalmonte2011,Alcaraz2011,Alcaraz2012,Dalmonte2012}.
These results support the following  scenario for the models considered here: 
while  the XXZ model 
exhibits oscillatory corrections to the CFT formula,
the discrete-symmetric $Z_2$ transverse field Ising and $Z_3$ three-state Potts models
exhibit no oscillations.
Furthermore, the power law decay exponent of the leading term of $C_{\alpha}(L)$  is given by 
\begin{equation}
p_{\alpha} =\frac{\tilde{\eta} X_e}{\alpha},
\end{equation}
where $\tilde{\eta} = 1,2$ for OBC/PBC, and $X_e$ is the scaling dimension of the energy operator.
An exception is  the von Neumann entropy of systems with PBC.
In this case, the leading term of $C_{\alpha}$ 
does not oscillates with $L$, and the power law decay is given by $1/L^{\nu}$, 
where $\nu = 2$, as shown by numerical  results based on DMRG \cite{Alcaraz2012}. 

It is worth mentioning that corrections with scaling exponent $p_{\alpha} = 2X_e/\alpha$ are related to the 
appearance of relevant operators at the conical singularities in the CFT geometry of the path integral representation of the reduced density matrix\cite{Cardy2010}. Due to bulk irrelevant operators there are also some unusual corrections of the 
form $L^{-2(x-2)}$ and $L^{2-x-\frac{x}{\alpha}}$ for the values of $\alpha$ close to one and $\alpha>\frac{x}{x-2}$, respectively\cite{Cardy2010}. We will not 
investigate these extra subleading corrections in this work.

\subsection{Diagnostics for the accuracy of the Bisognano-Wichmann entanglement Hamiltonian on the lattice}
\label{accuracyBW}

The accuracy of $\rho_{BW}$ relies on the underlying field theory being Lorentz
invariant~\footnote{For the BW-EH  obtained from Eq.\eqref{infinitePBC},\eqref{finiteOBC},\eqref{finitePBC}, the accuracy of $\rho_{BW}$ relies on the underlying field theory being conformal invariant}. 
This is always the case for the quantum critical chains considered here, where conformal symmetry emerges as a feature of the low-energy degrees of freedom of the system.
Even in this case, however, one shall expect that lattice effects are not completely suppressed, and the exact EH is not exactly given by Eq.~\eqref{EHBW}.
As an example, we mention the exact results for the EH of a free fermionic chain at half-filling,
that are very close to the BW-EH, but presents tiny longer range terms that survives even in the $L \to \infty$ limit\cite{peschel2009,Peschel2018}.
These terms, completely absent in the BW-EH, are caused by the curvature arising in the dispersion relation of hopping fermions on the lattice away from Fermi points and they can be systematically computed~\cite{Tonni2019}.
In the context of lattice models the exact EH for a half chain is known 
for the transverse field Ising chain away from the criticality and the XXZ model in the massive phase \cite{Itoyama:1987aa,Peschel1999}.

Since even when $L \to \infty$ the BW-EH is not exact in general (e.g., as for the free-fermion chain),
one may wonder that the BW R\'enyi entropy: 
(i) does not reproduce the non-universal contributions such as the additive constants   $c_{\alpha}^{\prime}$  (or $c_{\alpha}^{\prime\prime}$),
despite the fact that the low energy part of the spectrum of the BW-EH is in almost perfect agreement with the exact ones, as discussed in Ref.~\cite{Giuliano2018};
(ii) does not capture the corrections to the CFT scaling associated to relevant (or irrelevant) operators, as discussed in the last section.

In order to investigate these issues,  we consider  the size scaling of the R\'enyi entropy  
obtained from $\rho_{BW}$ in Sec. \ref{resultsI}.
More specifically, we discuss  the behavior of the leading terms of $S_{\alpha}^{BW}$ and
the BW corrections to the CFT formula
\begin{align}
 C_{\alpha}^{BW}(L) = S_{\alpha}^{BW}(L) - S_{\alpha}^{CFT}(L) + c^{\prime}_{\alpha},
 \label{eq:correction}
\end{align}
Note that for convenience, we add the constant $c^{\prime}_{\alpha}$ ($c^{\prime \prime}_{\alpha}$ for a finite system).
The asymptotic leading term of $C_{\alpha}^{BW}(L)$ is investigated by  fitting it  with the following function
\begin{align}
F_{\alpha}(L) = A_{\alpha} +  f_{\alpha}^{BW}/L^{p_{\alpha}^{BW}}, 
\label{fit}
\end{align}
where $A_{\alpha}$, $f_{\alpha}^{BW}$ and $p_{\alpha}^{BW}$ are free parameters.
As can be noted, apart from the constant $A_{\alpha}$, $F_{\alpha}(L)$, 
disregarding the oscillating factor, has the same form of  $d_{\alpha}(L)$ [Eq.\eqref{d}].
We also consider the discrepancy between the BW and the exact R\'enyi entropy
\begin{equation}
 dS_{\alpha} = |S_{\alpha}^{BW} - S_{\alpha}|.
\label{discr}
\end{equation}

In  Sec.~\ref{resultsII}, we turn our attention to the Schatten norm distance between the exact and the BW reduced density matrix, defined as:
\begin{align}
 N_p = (\tr {D}^p)^{1/p},
 \label{eq:Schatten}
\end{align}
where 
\begin{align}
 {D} = {\rho}_{BW} - {\rho}_{A},
 \label{Ndist}
\end{align}
and $p$ is an even integer number. 
Similarly to the Uhlmann fidelity and the trace norm \cite{Fagotti2013}, $N_p$ measures the
norm of the distance between
$\rho_{BW}$ and $\rho_{A}$. In the following, we focus on $N_p$, as it is easier to be computed numerically and is the only
quantity which can (at least in principle) be accessed in $D>1$ using quantum Monte Carlo methods.

We stress that, unlike previous studies that were mostly concerned with the low-lying part of the entanglement spectrum (ES), eigenvectors, and correlators~\cite{Dalmonte2017,Peschel2018,Kosior:2018aa,Giuliano2018,Turkeshi_2019}, we focus here on properties of the full reduced density matrix, 
such as momenta of the ES distribution (i.e., the REs) and properties that depend on all the eigenvectors (i.e., the norm distance). 
While the analysis of the size scaling of the REs allows us to check if $\rho_{BW}$  captures universal properties of the system, the norm distance probes if indeed $\rho_{BW}$ is able to reproduce accurately the actual state of the subsystem.

\section{Results I: BW R\'enyi entropy}
\label{resultsI}

In this section, we analyze the accuracy of  $S_{\alpha}^{BW}$ and $C_{\alpha}^{BW}$
by directly comparing it with the results obtained from the exact $\rho_{A}$, 
and the general theoretical behavior of $S_{\alpha}$ known from CFT.  
We consider the partitions shown in Fig.\ref{fig:cartoon}
(i.e., for finite systems we always consider half-partition).
In the next three subsections we discuss results for the XXZ chain,  
transverse field Ising (TFIM), three-state Potts (3SPM) and bilinear-biquadratic (BBM) models.
As anticipated in Sec.~\ref{renyiCFT}, the exact R\'enyi entropy 
exhibits an oscillatory behavior with respect to $L$ for the XXZ chain, 
whilst these oscillations are absent for both TFIM and 3SPM.
Finally, we conclude this section by showing   $S_{\alpha}^{BW}$ for the critical non-integrable BBM.

\subsection {XXZ model}

The XXZ model is defined as
\begin{align}
 H_{XXZ} = J \sum_{n=1}^{L_T}  S_{n}^{x}S_{n+1}^{x} + S_{n}^{y}S_{n+1}^{y} +\Delta S_{n}^{z}S_{n+1}^{z},
 \label{modelXX}
\end{align}
where $S_n^{i} = (1/2) \sigma_n^{i}$ are spin-1/2 operators, with $i=x,y,z$, and $\sigma_n^{i}$ are the Pauli matrices; the exchange coupling, $J$, sets the energy scale.
Here we focus on the parameter region $ -1 < \Delta \le 1 $, where the ground state of the XXZ is gapless and  can be described  by a CFT  with $c=1$.
In this regime, the exact sound velocity is given by~\cite{MussardoBook}
\begin{equation}
v = \frac{\pi\sqrt{1-\Delta^2}}{2\arccos\Delta}. 
\end{equation}
It is worth mentioning that the exact EH in the massive phase (i.e., $\Delta < -1$ or $\Delta > 1$)  is equal to the BW-EH [with coupling given by Eq. \eqref{infiniteOBC}] \cite{Itoyama:1987aa,Peschel1999} for $L\to\infty$.
However, the corner transfer matrix method used to obtain this result is not applicable to the gapless regime discussed here.

The resulting BW-EH for $\Delta =0$, that corresponds to the XX model,
can be mapped to a free-fermion Hamiltonian with the aid of the Jordan-Wigner transformation, see Appendix \ref{appendixA}.
The $S_{\alpha}^{BW}$ [Eq.\eqref{BWrenyi}] is then obtained by diagonalizing the $L \times L$ matrix.
This method allows us to achieve  very large subsystem sizes ($L \sim 10^{4}$),
which is fundamental to determine the corrections to the leading term in $S_{\alpha}^{BW}$. For $\Delta \neq 0$,  the  calculation of   $S_{\alpha}^{BW}$ is
limited  to $L \le 100$, and is performed using QMC and ED methods (see below).

\begin{figure}[]
{\centering\resizebox*{7.8cm}{!}{\includegraphics*{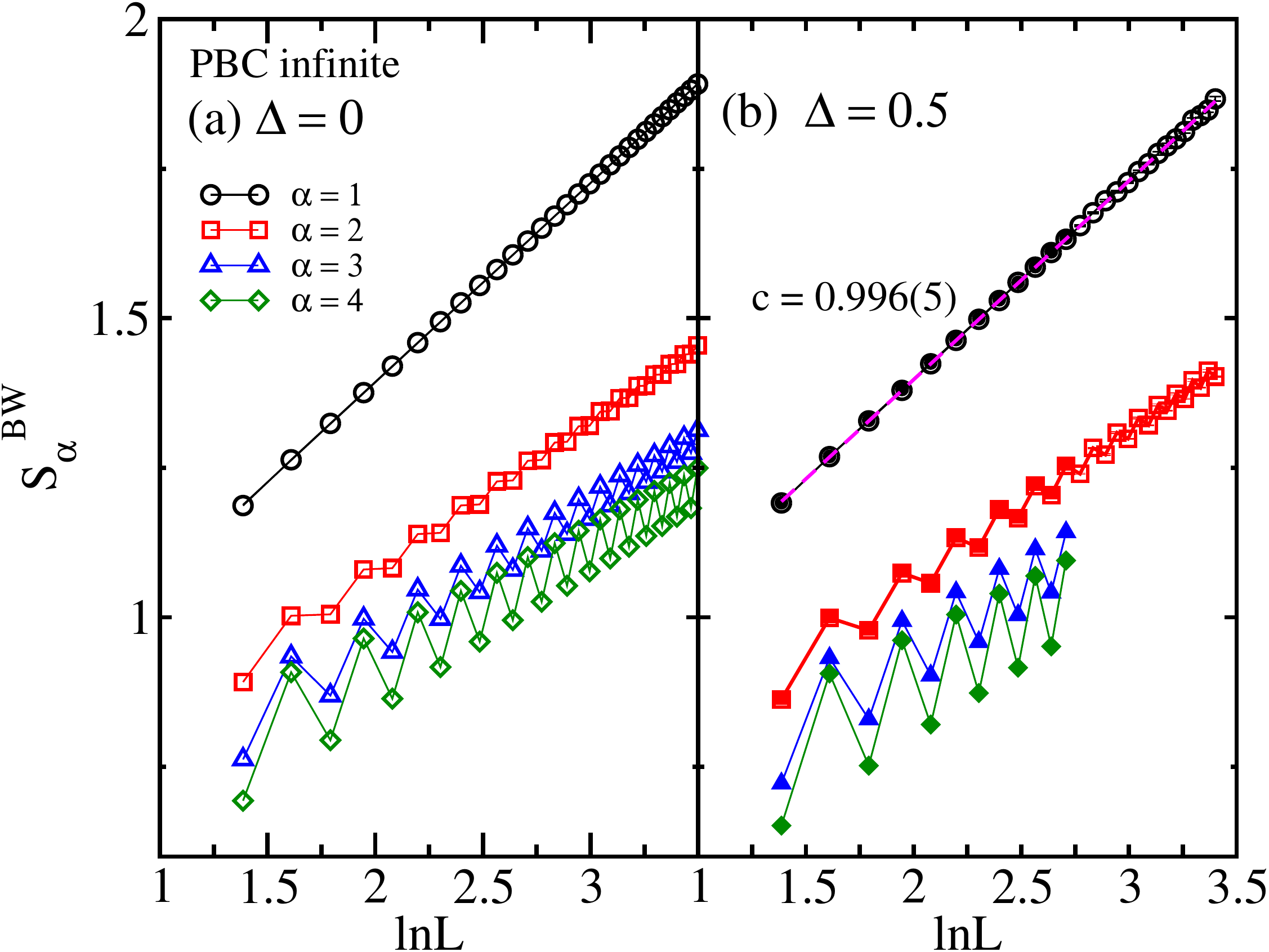}}}
\caption{BW R\'enyi entropy for the XXZ model with (a) $\Delta = 0$ and (b) $\Delta = 0.5$. Results for both even and odd values of $L$ are shown. 
In panel (a) we obtain $S_{\alpha}^{BW}$ by diagonalizing the corresponding free-fermion BW-EH of the XX model, while in
panel (b) we use both QMC (empty points) and ED (filled points). QMC error bars are smaller than the point sizes.
The values of the central charge extracted from $S_{1}^{BW}$ is presented in the panel (b).
}
\label{fig:entropy} 
\end{figure}

For the PBC case, the size-scaling of the  BW von Neumann entropy follows the expected behavior 
predicted by CFT, as discussed in Refs.~\onlinecite{peschel2009,Peschel2018,Mendes2019}, and illustrated in Fig.~\ref{fig:entropy}a for the $\Delta=0$ case.
We confirm this result for the XXZ model with $\Delta = 0.5$ using the Wang-Landau SSE method \cite{Wessel2003,Wessel2007}.
We consider the following cutoff for the SSE series expansion: $\Lambda~=~2.5\beta_{EH}|E(\beta_{EH})|$, where $E(\beta_{EH})$ is the expectation value of the total energy at 
inverse temperature $\beta_{EH}$. This choice of $\Lambda$ allows us to obtain $S_{\alpha}^{BW}$ for $\alpha \le 2$;
for comparison, we also compute $S_{\alpha}^{BW}$ with ED.
As shown in Fig. \ref{fig:entropy} (b), we obtain $c = 0.996(5)$ by fitting the QMC data, $S_{1}^{BW}(L)$, for $L \le 30$ with the CFT formula [Eq.~\eqref{Scft}].
More interestingly, Fig. \ref{fig:entropy} shows that, while  $S_{1}^{BW}$ is a 
smooth function, $S_{\alpha}^{BW}$ (for $\alpha > 1$) exhibits oscillations with $L$,
as expected for the exact results when $\alpha > 1$.
Furthermore, the decrease of these oscillations with $L$ suggests that $S_{\alpha}^{BW}$ recovers the CFT formula in the asymptotic regime.
Similar corrections to the CFT formula are observed in the OBC case \cite{Mendes2019}.
We now investigate in more detail these corrections by presenting the results for $C^{BW}_{\alpha}(L)$ for different partitions.

\begin{figure}[]
{\centering\resizebox*{8.5cm}{!}{\includegraphics*{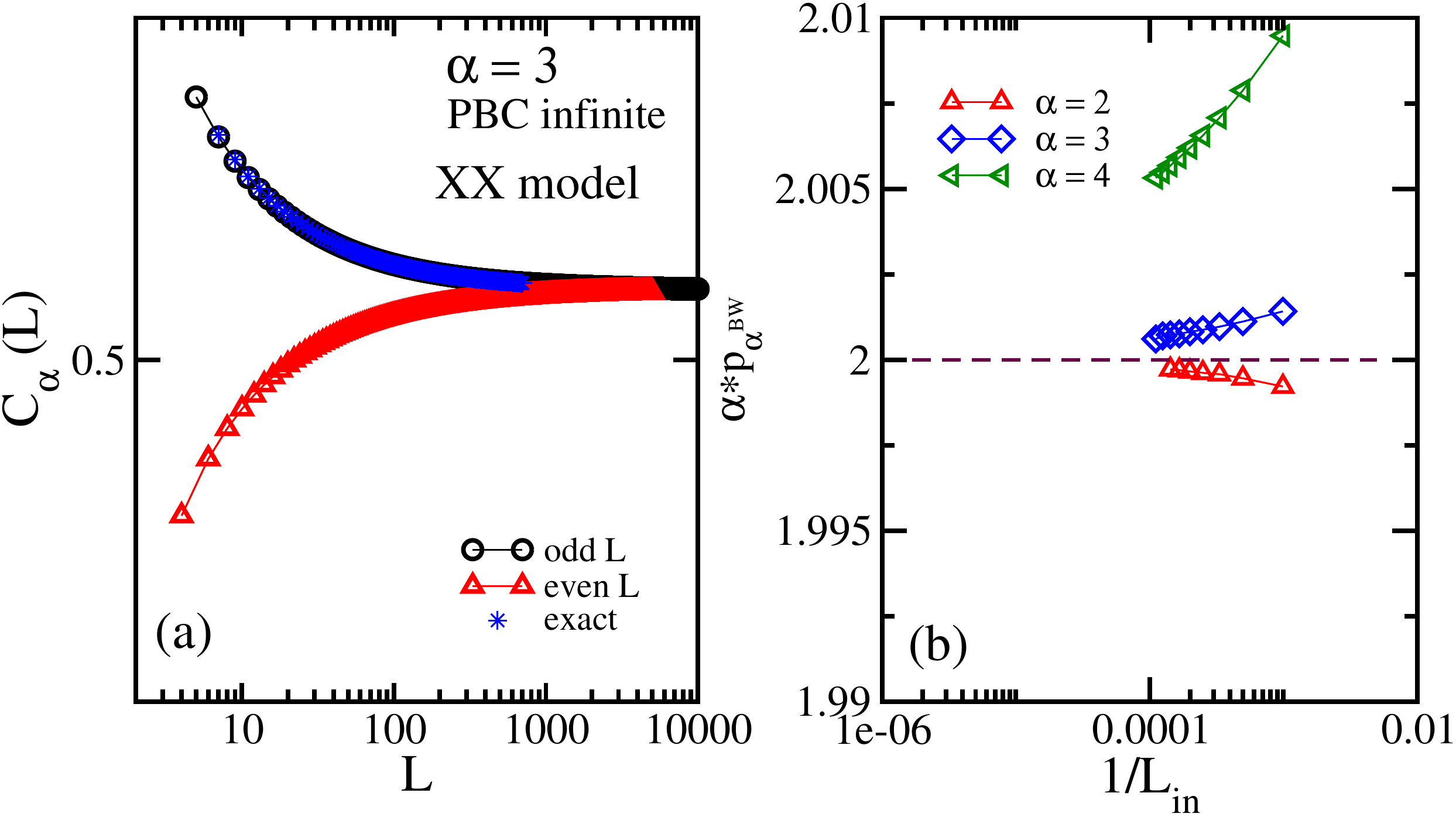}}}
\caption{\textit{Correction to the scaling of the BW R\'enyi entropy for an infinite system in the XX model.}
$C_{\alpha}$ is presented in panel (a) for even and odd values of $L$;
the blue points (stars) correspond to the exact results obtained with the correlation matrix technique \cite{Peschel2004}.
In panel (b) we show  $p_{\alpha}^{BW}$ calculated using 
the fitting procedure described in the main text;
the dashed line represents the exact value of $\alpha * p_{\alpha}$.}
\label{fig:infiniteXX} 
\end{figure}

\begin{figure}[]
{\centering\resizebox*{8.5cm}{!}{\includegraphics*{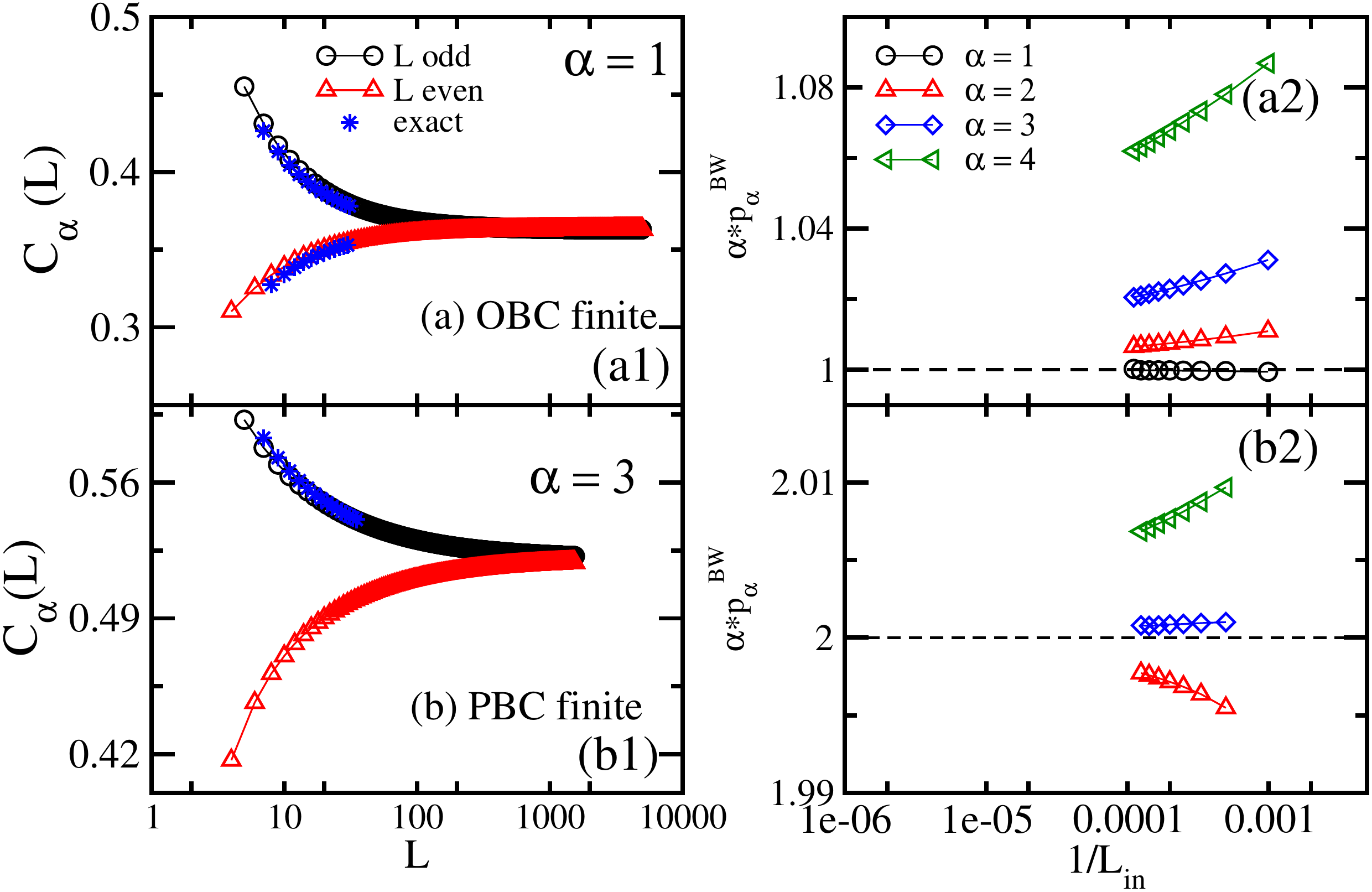}}}
\caption{\textit{Correction to the scaling of the BW R\'enyi entropy for the finite-system partitions in the XX model.}
Panels (a1) and (b1) present  $C_{\alpha}$ for  OBC and PBC, respectively; the blue points (stars)
correspond to exact results obtained with the correlation matrix technique \cite{Peschel2004}.
In panels (a2) and (b2) we show the results of the calculated $p_{\alpha}^{BW}$ for OBC and PBC, respectively, using 
the fitting procedure described in the main text; the dashed line represents the exact value of $\alpha * p_{\alpha}$.
}
\label{fig:finite} 
\end{figure}

\begin{table}
\centering
\begin{tabular}{ |c|c|c|c| } 
 \hline
       & $\alpha = 2$ & $\alpha = 3$ & $\alpha = 4$ \\ 
 \hline
 BW    & 0.11422(5) & 0.1614(5) & 0.1743(5) \\ 
 \hline
 exact & 0.11423... & 0.1609... & 0.1726... \\ 
 \hline
 \end{tabular}	
 \caption{The table shows the comparison between the calculated $f_{\alpha}^{BW}$ (using the infinite PBC BW-EH) for the XX model and the exact coefficients of Eq.\eqref{d}, $f_{\alpha}$,
 extracted from Ref.~\onlinecite{Calabrese2010_2}.}
 \label{table1}
\end{table}

\subsubsection{XX model ($\Delta  = 0$)}

We  first consider  $C_{\alpha}^{BW}(L)$  for the XX model in  the infinite PBC case.
The oscillatory behavior of $S_{\alpha}^{BW}$ is manifested when we plot $C_{\alpha}^{BW}(L)$  for even and odd values of $L$.
As an example we show the $\alpha = 3$ case in Fig. \ref{fig:infiniteXX} (a).
The asymptotic behavior of $C_{\alpha}^{BW}(L)$  is analyzed by fitting our results with Eq. \eqref{fit} (in the fit, we just consider the values of $C_{\alpha}^{BW}(L)$  for odd $L$).
Although not shown here, we also considered other values of $\alpha$, with similar conclusions.
The fits are performed with respect to $L$ in the interval $\left[L_{in},L_{max}\right]$;
where the maximum value considered for  $L_{in}$ is  $L_{in} = L_{max} - 10^{3}$.
As we improve the quality of the fit (i.e., increasing the value of $L_{in}$),
the parameter $p_{\alpha}^{BW}$ converges to the expected scaling exponent, $p_{\alpha} = 2/\alpha$, see Fig. \ref{fig:infiniteXX} (b).

It is worth mentioning that the coefficient $f_{\alpha}^{BW}$ obtained from the fit is also in a quantitative agreement with the exact result
of $f_{\alpha}$ calculated in Ref.~\onlinecite{Calabrese2010,Calabrese2010_2}, see Table \ref{table1}.
In the Table we considered $\alpha = 2,3$ and $4$;
the agreement of $f_{\alpha}^{BW}$  becomes worst for larger $\alpha$ because sub-leading terms of $C_{\alpha}^{BW}$ [in addition to the leading one described by Eq. \eqref{d}] 
become more relevant as we increase $\alpha$, as occur with the exact Renyi entropy \cite{Calabrese2010}. 
Note that for $\alpha = 1$ the exact calculations predict $f_1 = 0$. The  $f_1^{BW}$ will be discussed below.
These results  strongly indicates that, in the  asymptotic regime, $C_{\alpha}^{BW}(L)$ is not just qualitatively, but also quantitatively in agreement with
the leading asymptotic behavior of the exact corrections.

We now consider the finite-system partitions with both OBC and PBC; see Fig. \ref{fig:cartoon} (b) and (c).
The correction $C_{\alpha}^{BW}(L)$ exhibits the expected oscillatory behavior with $L$.
For OBC, this behavior occurs even for the $\alpha = 1$ case, see Fig. \ref{fig:finite} (b1).
Furthermore, by fitting these results with Eq. \eqref{fit}, we observe that the values of $p_{\alpha}^{BW}$ 
are also in agreement with the exact results $p_{\alpha} = 1/ \alpha$ and $p_{\alpha} = 2/\alpha$ for the OBC and  PBC, respectively, see Fig. \ref{fig:finite} (a2) and (b2). These results
indicate that the leading asymptotic term of $C_{\alpha}^{BW}$  is given by $d_{\alpha}$ [Eq. \eqref{d}].

Despite the agreement between $C_{\alpha}^{BW}$ and the exact results in the asymptotic limit, 
a comparison between  $C_{\alpha}^{BW}(L)$ and the exact results still  
shows some tiny discrepancies, see Fig. \ref{fig:finite}.
Although not visible in Fig. \ref{fig:infiniteXX} (a) these tiny discrepancies also exist for the infinite PBC case.
The question then is what is the nature of these discrepancies.
The results discussed so far indicate the following: while the leading term of  $C_{\alpha}^{BW}(L)$ 
coincides with $d_{\alpha}$  [Eq.\eqref{d}]
subleading corrections, that are most likely present in both the exact and the BW $C_{\alpha}(L)$, are different (at least, at the scale accessible to our numerical calculations).

\begin{figure}[]
{\centering\resizebox*{8.5cm}{!}{\includegraphics*{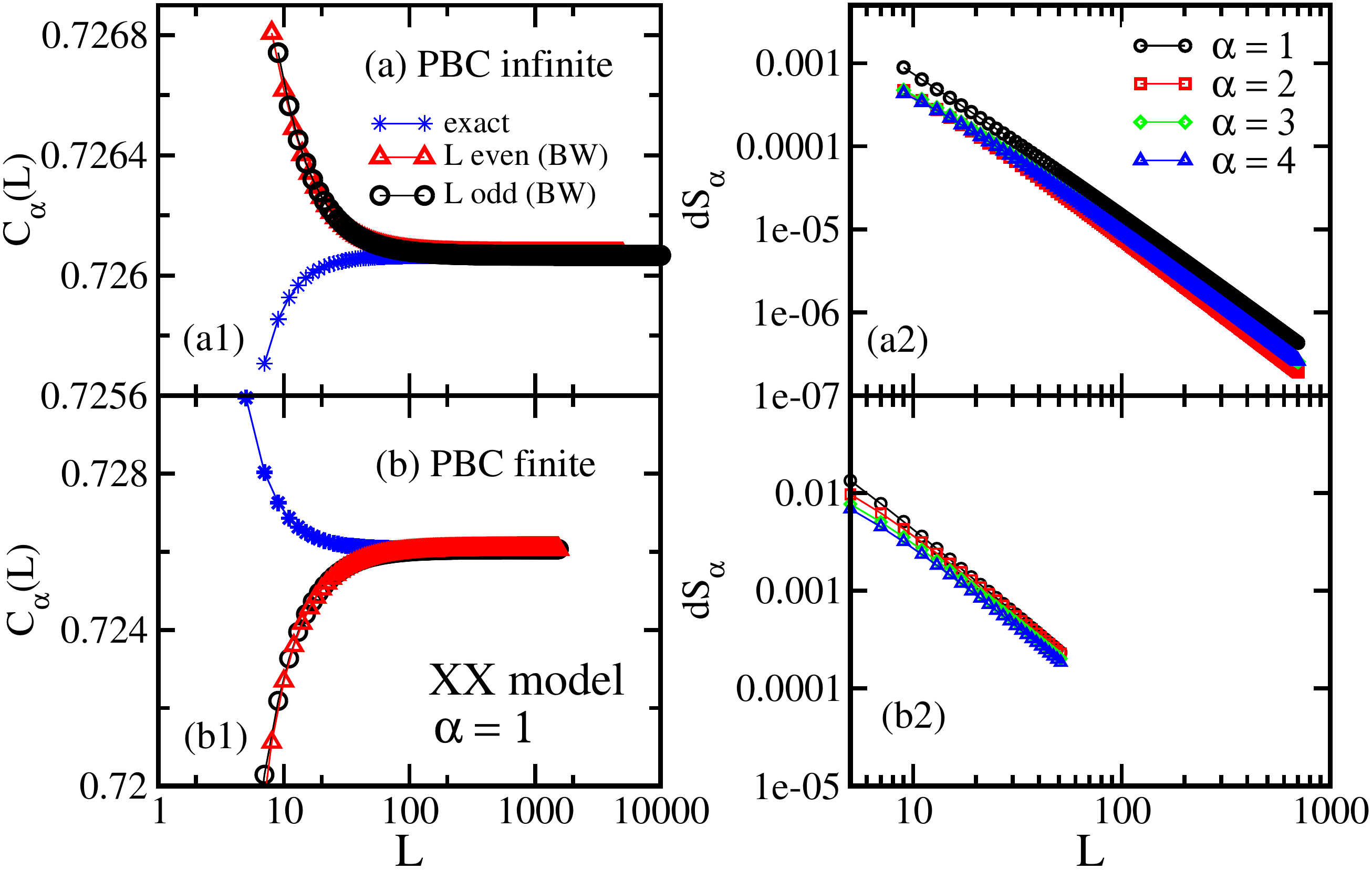}}}
\caption{Panels (a1) and (b1) present the corrections to the scaling of the BW R\'enyi entropy, $S_{1}^{BW}$, for the infinite and finite PBC partitions, respectively, in the XX model. 
The blue points correspond to exact results.
In panels (a2) and (b2) we show the discrepancy between the BW and the exact REs  for 
the infinite and the finite PBC partitions, respectively, with different values of $\alpha$}
\label{fig:dS} 
\end{figure}

\begin{figure*}[]
{\centering\resizebox*{8.5cm}{!}{\includegraphics*{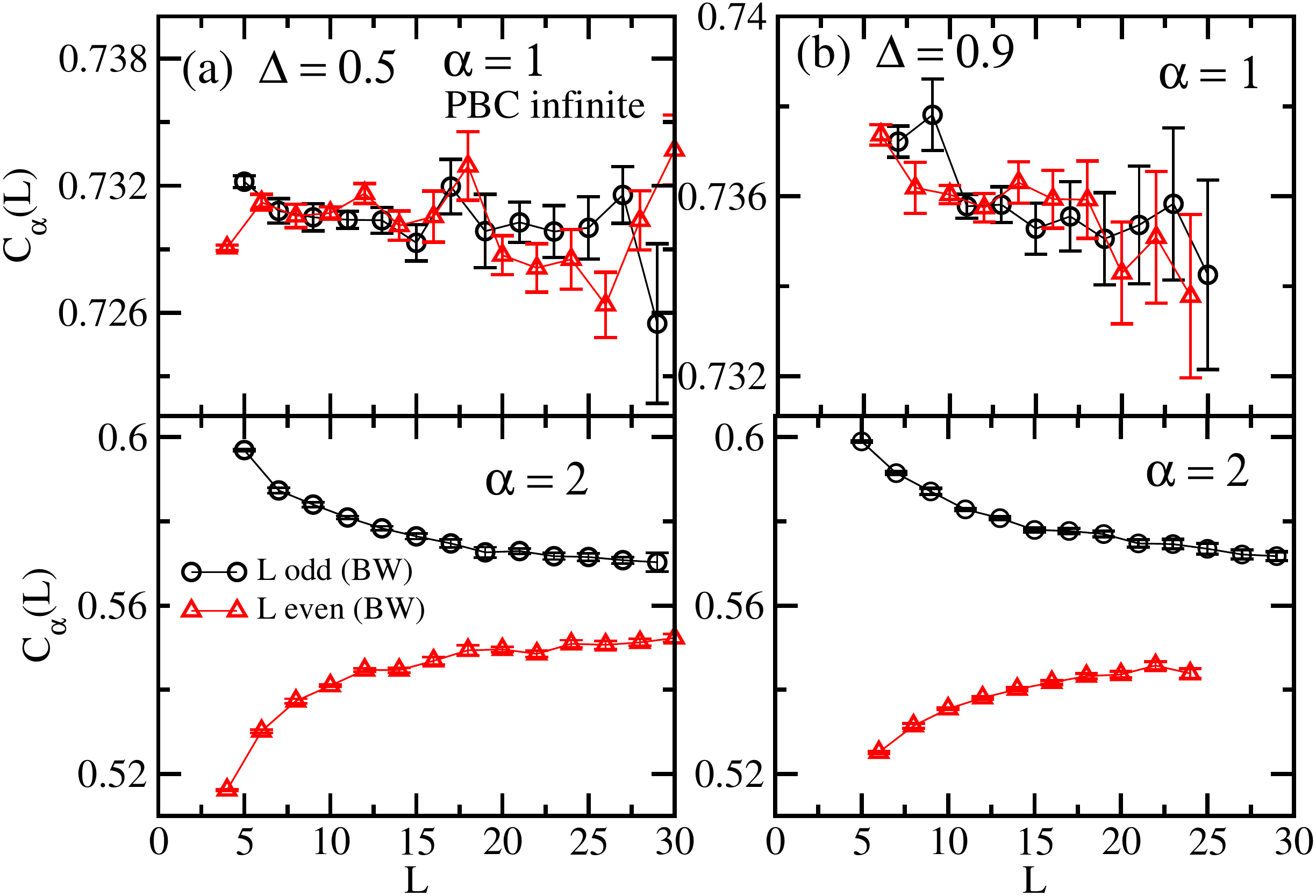}}}
{\centering\resizebox*{7.75cm}{!}{\includegraphics*{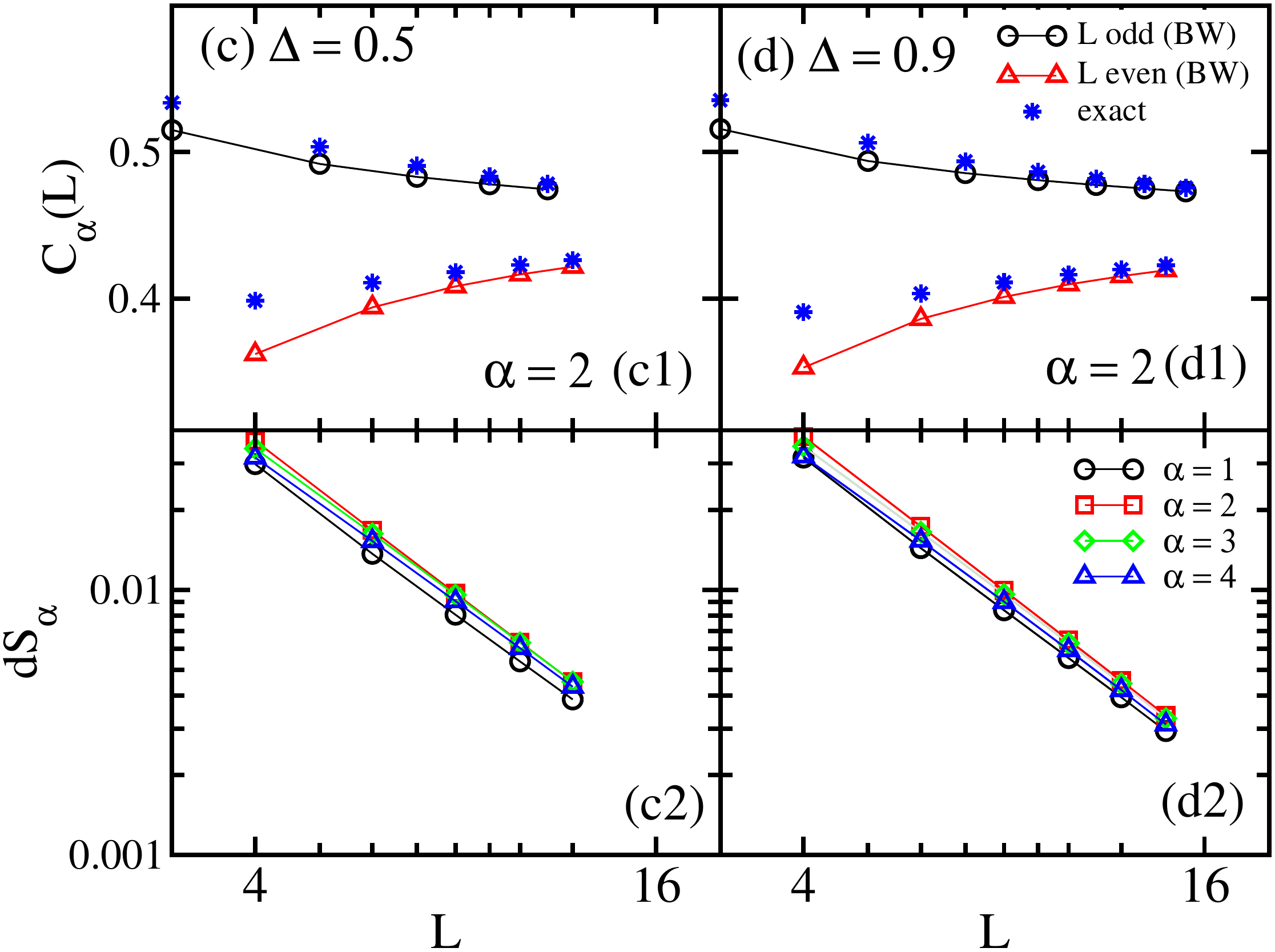}}}
\caption{\textit{BW R\'enyi entropy in the XXZ model for $\Delta \neq 0$.} 
In panels (a) $\Delta = 0.5$ and (b) $\Delta = 0.9$ we present the correction to the scaling of $S_{\alpha}^{BW}$ obtained with QMC for the infinite PBC partition.
In panels (c) $\Delta = 0.5$ and (d) $\Delta = 0.9$ we obtain the results with ED and DMRG (see text) for the finite PBC partition: (c1) and (d1) show the corrections to the scaling of $S_{2}^{BW}$, while in
(c2) and (d2) we  present the discrepancy between the BW and the exact REs for different values of $\alpha$.
}
\label{fig:XXZ} 
\end{figure*}

In order to better analyze the point raised in the last paragraph, 
we discuss now the behavior of the von Neumann entropy for systems with PBC.
In this case, the exact $S_{1}$ does not exhibit any oscillating term, 
and the leading term of $C_{1}$ is not given by $d_{\alpha}$ \cite{Calabrese2010,Calabrese2010_2}.
It is interesting to note that $S_1^{BW}$ also does not oscillate with $L$,
as can be seen in Fig. \ref{fig:dS}.
Nevertheless, unlike the $\alpha > 1$ cases, the trends of the size scaling of the BW and the exact  $C_{\alpha}$ are completely different,
which explains the tiny discrepancy, i.e., $dS_{\alpha}/S_{\alpha} < 10^{-3}$.
In this case, already the leading term of $C_{\alpha}^{BW}(L)$ differs from the exact corrections.

We note that the scaling exponent that determines the asymptotic behavior of $dS_{\alpha} \sim 1/L^{\gamma_{\alpha}}$
does not depend on  $\alpha$, and is $\gamma_{\alpha} \approx 1.9$, see Fig. \ref{fig:dS} (a2) and (b2).
The fact that $\gamma_{\alpha}$ is independent of ${\alpha}$ is in line with the previous statement that the correction $d_{\alpha}$ is present in both the 
BW and the exact $S_{\alpha}$, i.e., this factor cancels out when we consider the difference $dS_{\alpha}$.
Finally, we observe  that the discrepancy of the BW REs is almost independent of  $\alpha$, as expected from the discussion above.

\subsubsection{$\Delta  \neq 0$}

We now discuss the results for $\Delta \neq 0$.
In this case, the investigation of the asymptotic behavior of $C_{\alpha}^{BW}$ is hindered by the small system sizes that one can achieve
with ED.
In addition, the tiny discrepancies between the BW and the exact results, $dS_{\alpha}/S_{\alpha} < 0.1\%$, are difficult to access with QMC due to the
statistical errors in the MC estimates.
Despite these technical issues, the results obtained with both ED and QMC show that the behavior of  $C_{\alpha}^{BW}$ is in line with the exact results.
The exact R\`enyi entropy (and $C_{\alpha}$) is obtained with ED for subsystem sizes $L \le 12$ and DMRG for $  L > 12$.
In the DMRG calculation we obtain the entanglement
spectrum of the original system by keeping
$100$ - $150$ states and using the ground state as the target state in the proper symmetry sector.

In Fig.~\ref{fig:XXZ} (a) and (b), we show some examples of the scaling of $C_{\alpha}^{BW}$ obtained with QMC for $L \le 30$, and two different values of $\Delta$.
For PBC, $C_{\alpha}^{BW}$  oscillates  with $L$ for $\alpha = 2$, but not for $\alpha = 1$, as it is expected for the exact $C_{\alpha}$.
By fitting $C_{2}^{BW}$ with Eq.~\eqref{fit}, we obtain the following values for the scaling exponent $p_{2}^{BW}$ for two case values of the anisotropy $\Delta$:
$p_{2}^{BW} (\Delta=0.5)=0.78(3)$  and  $p_{2}^{BW}(\Delta=0.9)=0.61(7)$.
These results  have a discrepancy of almost $4\%$ with respect to the the exact results:
$p_{2}=0.75$ ($\Delta=0.5$) and $p_{2}\approx0.583$ ($\Delta=0.9$),
for the XXZ model with PBC, $p_2 = K$, where $K = \pi /\left( 2 \arccos (-\Delta) \right)$ is the Luttinger liquid parameter.
This discrepancy seems to be unaffected by potential logarithmic corrections that are present at  the isotropic point.

The results of $C_{\alpha}^{BW}$ obtained with ED for $L \le 15$ are also in agreement with the exact ones, as can be seen in Figs. \ref{fig:XXZ} (c) and (d).
The discrepancy $dS_{\alpha}$ goes to zero as a power law, $dS_{\alpha} \sim 1/L^{\gamma_{\alpha}}$, see Figs.~\ref{fig:XXZ} (c2) and (d2).
Furthermore, the scaling exponent $\gamma_{\alpha}$ is almost independent of $\alpha$, 
as observed for the XX model with larger subsystem sizes.
This feature can be explained if we assume that the correction $d_{\alpha}$ is present in both the 
BW and the exact $S_{\alpha}$, i.e., this factor cancels out when we consider the difference $dS_{\alpha}$.

\subsection{Transverse field Ising and quantum three-states Potts models}

\begin{figure}[]
{\centering\resizebox*{7.8cm}{!}{\includegraphics*{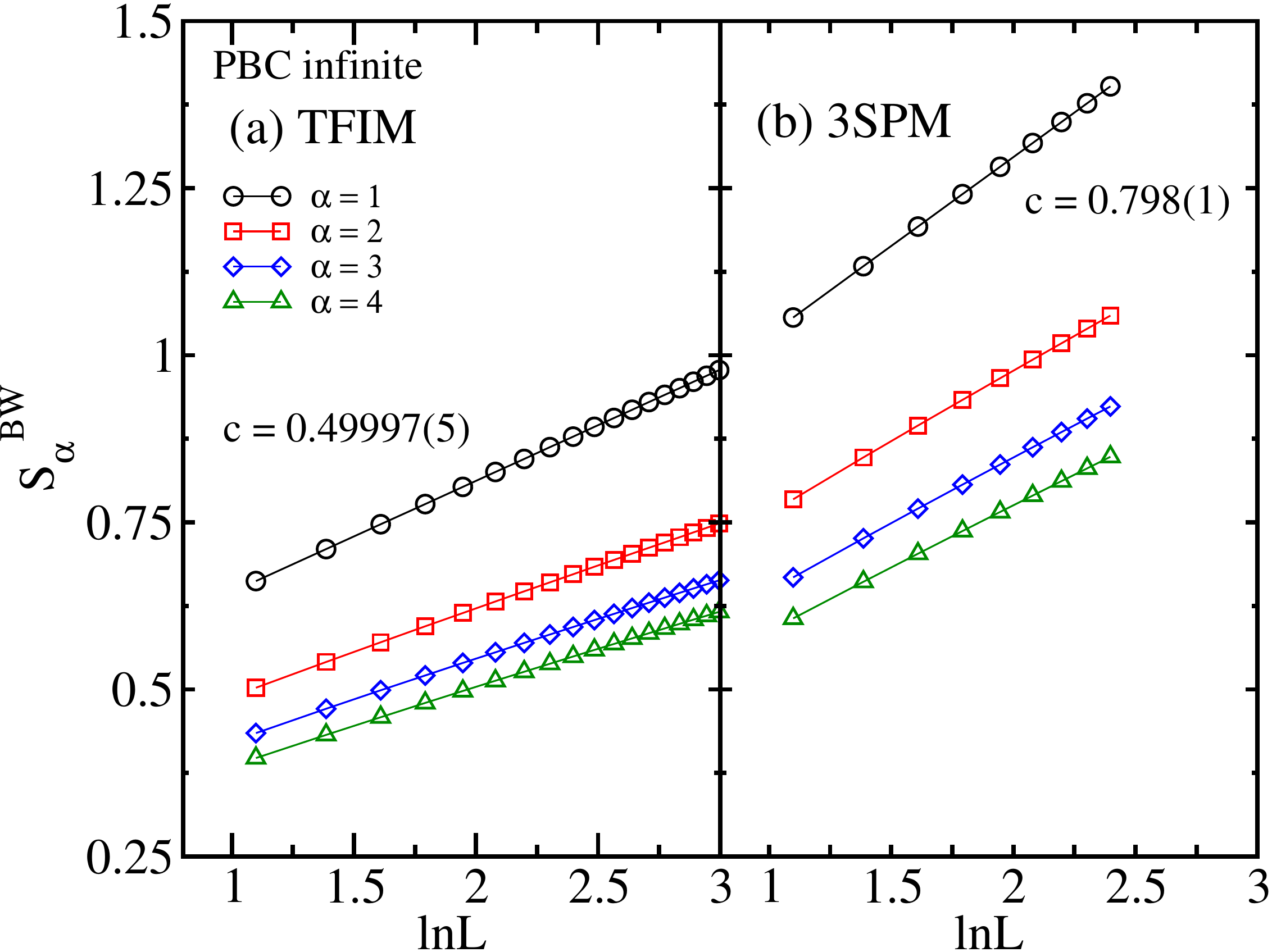}}}
\caption{BW R\'enyi entropy for the  (a) transverse field Ising 
and the (b) three-state Potts model for different values of $\alpha$. Results for both even and odd values of $L$ are shown.
In panel (a) we obtain $S_{\alpha}^{BW}$ by diagonalizing the corresponding free-fermion BW-EH of the TFIM, while in
panel (b) we use ED (see text).
The values of the central charge extracted from $S_{1}^{BW}$ are presented in the panels. }
\label{fig:entropyZn} 
\end{figure}

\begin{figure}[]
{\centering\resizebox*{8.5cm}{!}{\includegraphics*{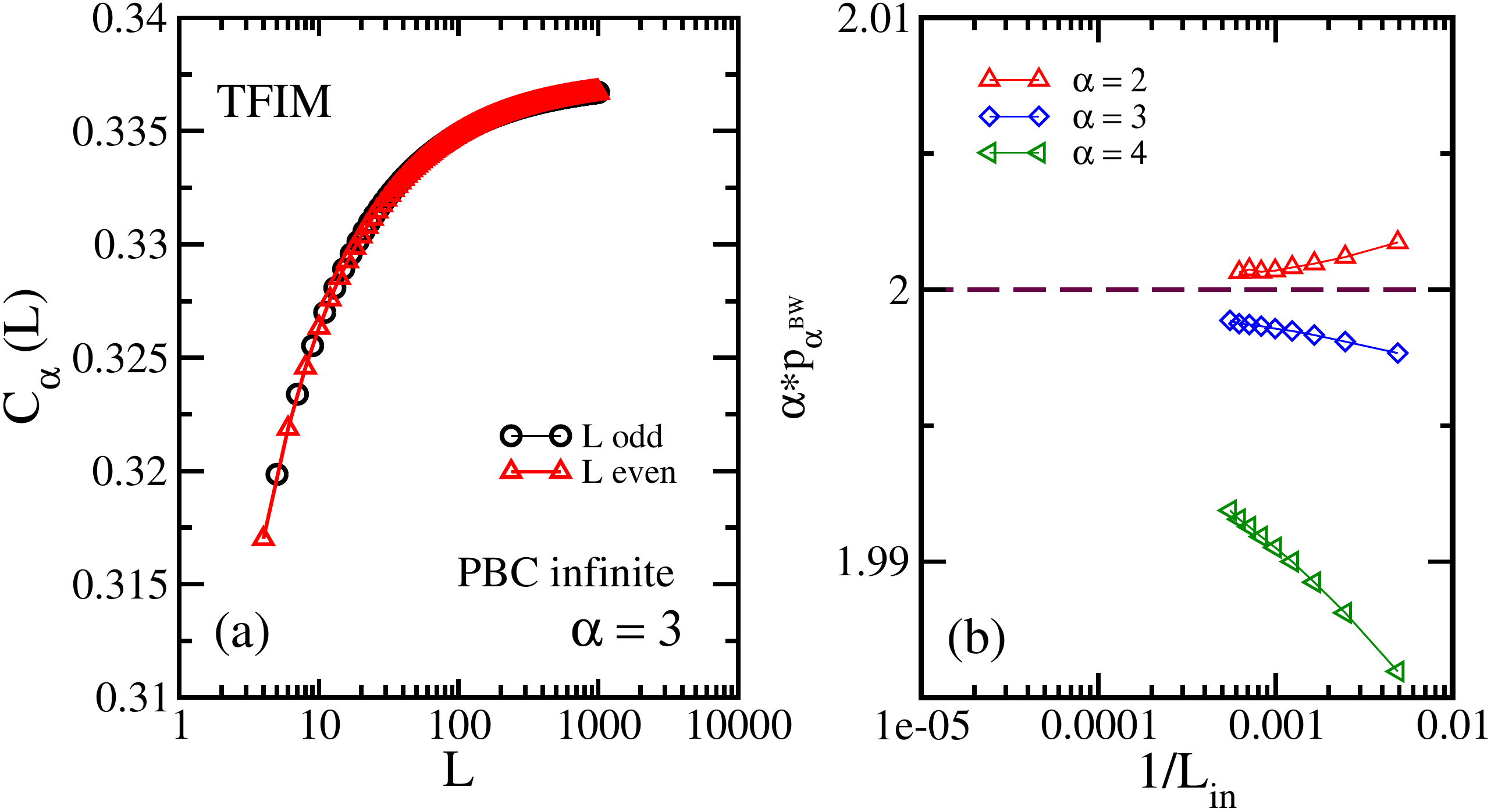}}}
\caption{\textit{Correction to the scaling of the BW R\'enyi entropy for an infinite system in the transverse field Ising model.} 
$C_{\alpha}$ is presented in panel (a), while in panel
(b) we show the results of the calculated $p_{\alpha}^{BW}$ using 
the fitting procedure described in the main text; the dashed line represents the exact value of $\alpha * p_{\alpha}$.}
\label{fig:inifiniteIsing} 
\end{figure}

In this section, we discuss the behavior of $S_{\alpha}^{BW}$ for models that are characterized by discrete global symmetries.

First, we consider the transverse field Ising model(TFIM)~\cite{fradkinbook}:
\begin{align}
 H_{TFIM} = - \sum_{n=1}^{L_T} S_{n}^{z}S_{n+1}^{z} - g \sum_{n=1}^{L_T} S_{n}^{x}.
 \label{modelIsing}
\end{align}
This model displays a quantum phase transition between a ferromagnetic and a disordered ground state at $g=1$. In the latter phase, a global $\mathbb{Z}_2$ symmetry corresponding to spin inversion is spontaneously broken.
It is worth to point out that for this model the EH of a half-partition   
in an infinite chain was computed exactly away from the critical point~\cite{peschel2009}. 
The result perfectly matches our lattice version of the BW theorem; however, it does not predict the prefactor $\beta_{EH}$. 
Here, instead, we focus at the quantum critical point, where the TFIM is characterized by a CFT with $c = 1/2$ and the exact sound velocity is equal to $v = 2$.
As it occurs for the XX model, the BW-EH of the TFIM can be mapped to a quadratic Hamiltonian with
the aid of the Jordan-Wigner transformation; see Appendix \ref{appendixA}.
For this model we are able to consider systems up to size $L = 10^{3}$.

We further consider the three-state Potts Model (3SPM)~\cite{sol-81,MussardoBook} 
\begin{align}
H=-\sum_{i=1}^{L_T} \left(\sigma_i\sigma_{i+1}^{\dagger}+\sigma_{i}^{\dagger}\sigma_{i+1}\right)-g\sum_{i=1}^{L_T} \left(\tau_{i}+\tau_{i}^{\dagger} \),
\label{H3pm} 
\end{align}
where $g>0$. The $\sigma$ and $\tau$ matrices are defined as
\be 
\sigma \ket{ \gamma } = \omega^{\gamma - 1 } \ket{\gamma},
\quad \quad 
\tau \ket{ \gamma } = \ket{ \gamma + 1 },
\vspace{2mm}
\quad \quad \omega = e^{i 2 \pi/3 },
\ee
and 
$\gamma = 0,1,2$. 
This model is characterized by a $\mathbb{Z}_3$ symmetry, which is spontaneously broken for $g < 1$ in the ferromagnetic phase.
As in the TFIM case, we focus on the quantum critical point, $g=1$. 
In this case the system is described by a CFT with the central charge $c=4/5$~\cite{dot-84,CFT1997},
and the exact sound velocity is $v= 3 \sqrt{3}/2$\cite{kedem1994}.
We use ED to obtain the $S_{\alpha}^{BW}(L)$ for systems with sizes up to $L = 12$. 

\begin{figure*}[]
{\centering\resizebox*{8.65cm}{!}{\includegraphics*{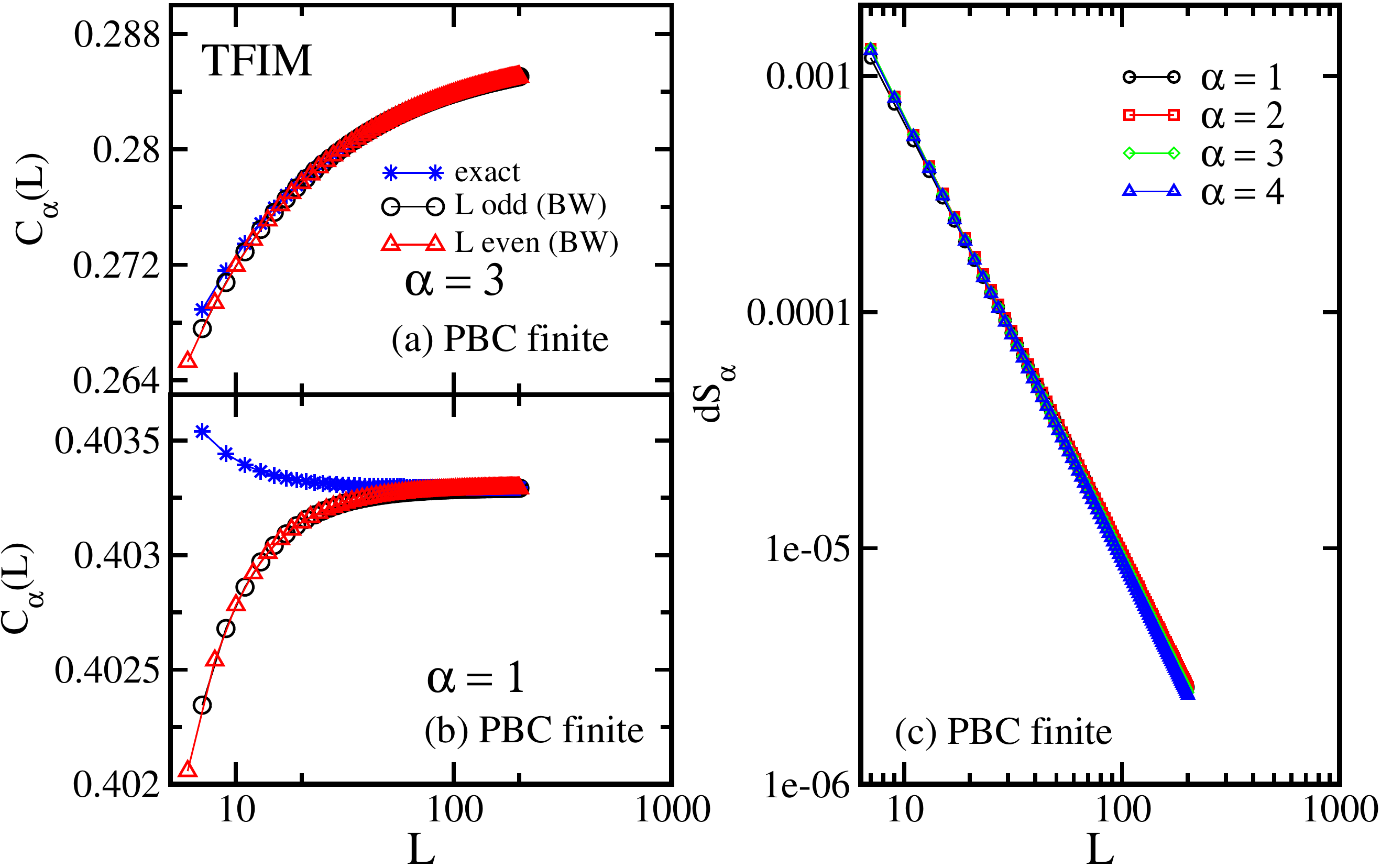}}} \ \ \ \ \ \ \
{\centering\resizebox*{8.45cm}{!}{\includegraphics*{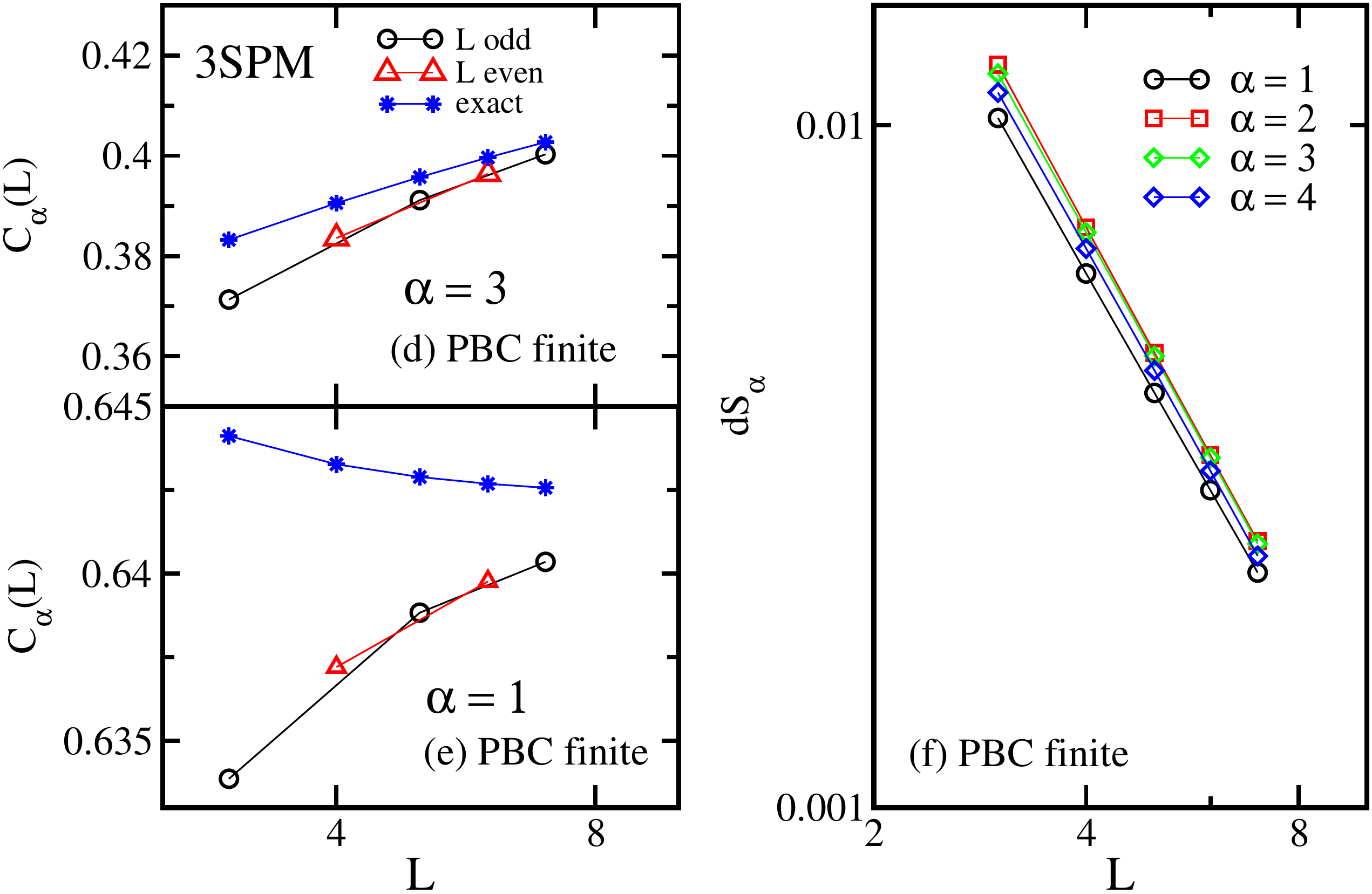}}}
\caption{\textit{Results for the transverse field Ising model are presented in panels (a)-(c)}.
Panels (a) and (b) show the correction to the scaling of the BW R\`enyi entropy for the finite PBC partition with $\alpha = 3$ and $\alpha = 1$, respectively,
while in panel (c) we consider the discrepancy, $dS_{\alpha}$, as a function of the subsystem size, $L$.
We obtain these results by diagonalizing the corresponding free-fermion BW-EH.
\textit{In panels (d)-(f) are presented results for the three-states Potts model} with the same set of parameters of panels (a)-(c), respectively, and
using ED (see text).}
\label{fig:dSZn}
\end{figure*}

Before discussing the scaling properties of the deviations with respect to the exact result, we briefly illustrate the overall scaling of $S_{\alpha}^{BW}(L)$ as a function of $L$. The latter is depicted in Fig.~\ref{fig:entropyZn}, for both the TFIM and the 3SPM. No oscillations as a function
with $L$ are present, as expected for these models~\cite{Alcaraz2012}.
Here we just show results for the infinite PBC case, however, we also confirmed similar results  for the other EHs described in the Sec. \ref{BWlattice}.
Furthermore, we calculated the central charge by fitting the  $S_{\alpha}^{BW}$ to the CFT formula, Eq.\eqref{Scft}. 
The outcome is in perfect agreement with the  exact results, as can be seen in Figs. \ref{fig:entropyZn} (a) and (b).
In particular, for the 3SPM we obtain $c \approx 0.798$, from $S_{1}^{BW}$, which has a discrepancy of just  $0.3\% $ with respect to the exact result, $c=4/5$.
It is worth emphasizing that this result is obtained for subsystem sizes up to just $L \le 12$,
which signals the fact that $S_{1}^{BW}$ is barely affected by corrections to the CFT formula, 
contrarily to what is observed for the R\'enyi entropy of the XXZ model.

\subsubsection{Transverse field Ising model}

We now discuss the behavior of the corrections to the CFT formula, $C_{\alpha}(L)$.
We start by considering an infinite system, and fit $C_{\alpha}^{BW}(L)$ for $L$ within the interval $[L_{in}, 10^{3}]$ for different values of $\alpha$ to Eq.\eqref{fit}, see Fig.~\ref{fig:inifiniteIsing} (a). 
As we increase $L_{in}$, the parameter $p_{\alpha}^{BW}$ converges to the exact result $p_{\alpha} = 2/\alpha$\, \cite{Calabrese2010} 
see Fig.~\ref{fig:inifiniteIsing} (b).

For the finite-system partitions, we explicitly compare $C_{\alpha}^{BW}$ with the exact results. In this case, we focus on the PBC case.
As we see in Fig.~\ref{fig:dSZn} (a), for $\alpha = 3$, there is a modest discrepancy between the 
exact and the BW results [e.g., $dS_{3}/S_{3} < 10^{-3}$ even for $L \approx 10$].
The exact and the BW corrections have the same behavior as $L$ increases, i.e., $C_{\alpha}^{BW}(L)$ increases and then saturates to
a constant value, which also strongly indicates that the leading term of the exact and the BW $C_{\alpha}(L)$ are the same. 
Consequently one can conclude that the discrepancy $dS_{\alpha}$ is due to subleading terms present in both the BW and the exact $C_{\alpha}(L)$.
Although not shown here, we have also observed similar results for all the $\alpha > 1$ cases that we considered.

For the case of the von Neumann entropy, the size scaling of $C_{\alpha}^{BW}$ has a different trend compared to the exact results, see Fig. \ref{fig:dSZn} (b).
As discussed for the XXZ model, since the exact $C_{1}$ differs from $d_{\alpha}$ [Eq. \eqref{d}] \cite{Calabrese2010,Calabrese2010_2,Alcaraz2012},
in this case, already the leading term of $C_{1}^{BW}$ is different from the exact corrections.

Finally, we observe that  $dS_{\alpha}$ goes to zero as a power law, $dS_{\alpha} \sim 1/L^{\gamma_{\alpha}}$, see Figs. \ref{fig:dSZn} (c).
The scaling exponent $\gamma_{\alpha}$
does not depend on  $\alpha$  ($\gamma_{\alpha} \approx 1.8$); see Fig. \ref{fig:dSZn} (c).
Similarly to the conclusions that we drew for the XXZ model, this result can be also explained by the fact that  the $\alpha$-dependent correction $d_{\alpha}$ [see Eq.A\eqref{d}] is present in both the 
BW and the exact $S_{\alpha}$ with comparable numerical coefficients.

\subsubsection{Three-states Potts model}

For the 3SPM, due to size limitation, we were not able to compute the exponent $p_{\alpha}^{BW}$.
Nevertheless, in Fig.~\ref{fig:dSZn} [panels (d)-(f)],  we show how the behavior of $C_{\alpha}^{BW}$ is qualitatively in agreement
with exact results.
In particular, we note that the scaling exponent $\gamma_{\alpha}$ associated to the power law, $dS_{\alpha} \sim 1/L^{\gamma_{\alpha}}$, see Fig. \ref{fig:dSZn} (f),
is almost independent of $\alpha$.

To summarize our investigation of the accuracy of the BW REs, we observed that
for all the models considered here
the almost negligible  discrepancies  (e.g., $dS_{\alpha}/S_{\alpha} < 10^{-3}$, even for subsystems with $L \approx 10$)  goes to zero as $L \to \infty$.
For a finite partition, the corresponding  value of $dS_{\alpha}$ can be understood as follows:
both the (i)  logarithmically-divergent CFT term, Eq. \eqref{Scft}, and
(ii) the leading corrections to the CFT scaling [see Eq.\eqref{d}] are properly described by $S_{\alpha}^{BW}$ (with the exception of $\alpha = 1$ for PBC);
and finally (iii) $dS_{\alpha}$ is related to subleading corrections, that are associated to both the BW and the exact results.
The investigation of the nature of the subleading corrections to $C_{\alpha}^{BW}$ is beyond the scope of this work.

\subsection{Bilinear-biquadratic model}

All the models under investigation so far can be either mapped into free-fermion chain or solved by Bethe ansatz,
and thus are considered integrable models.
In fact integrability is a key ingredient to carry out the analysis presented in the previous sections, because it gives the exact value of the sound velocity, $v$ in Eq.~\eqref{beta}, 
which allows the computation of $S_{\alpha}^{BW}$ [see Eqs.~\eqref{BWrenyi} and \eqref{BWvonneumann}].
However, as one can numerically compute  $v$ for non-integrable models,
it is worth to  consider the behavior of $S_{\alpha}^{BW}$ in these cases.
Thus, before closing this section, we consider the bilinear-biquadratic model (BBM)
\begin{equation}
H = \sum_{n=1}^{L_T} \vec{S}_n \cdot \vec{S}_{n+1} + g ( \vec{S}_n \cdot \vec{S}_{n+1} )^2.
\end{equation}
The phase diagram of this model hosts a gapped Haldane phase for $-1 < g < 1$. 
The two boundaries of this phase are gapless Bethe-integrable points whose underlying CFTs have central 
charges $c=3/2$ for $g=-1$ and $c=2$ for $g=1$ \cite{Alcaraz1988,itoi1997}. 
For $g<-1$ the system is gapped, while for $g>1$, although the integrability is lost, the low-energy degrees of freedom are described by a CFT with $c=2$. 
Here we consider the region $g>1$, and investigate the applicability of the BW-EH ansatz in the absence of integrability. 
We use periodic boundary conditions in order to exploit translation symmetry in the computation of the exact $S_{\alpha}$.
The system is frustrated when $L_T$ is not a multiple of $3$. 
As we need even system sizes in order to compute half-system entanglement, we use ED to compute $S_{\alpha}$ for  $L_T=6,12,18$.

\begin{figure}[]
\hspace{-7mm}
{\includegraphics[scale=0.42]{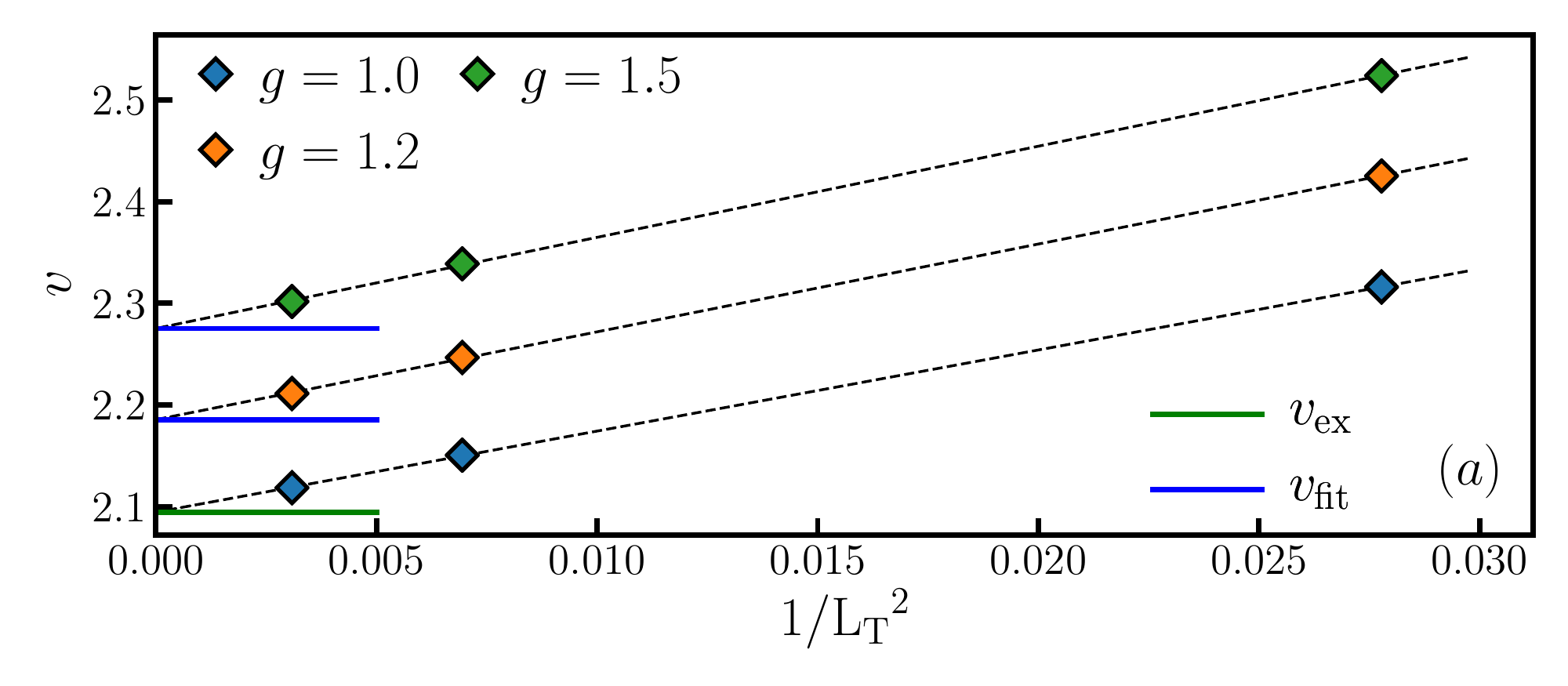} \\
\vspace{-1mm} \hspace{-7mm} \includegraphics[scale=0.44]{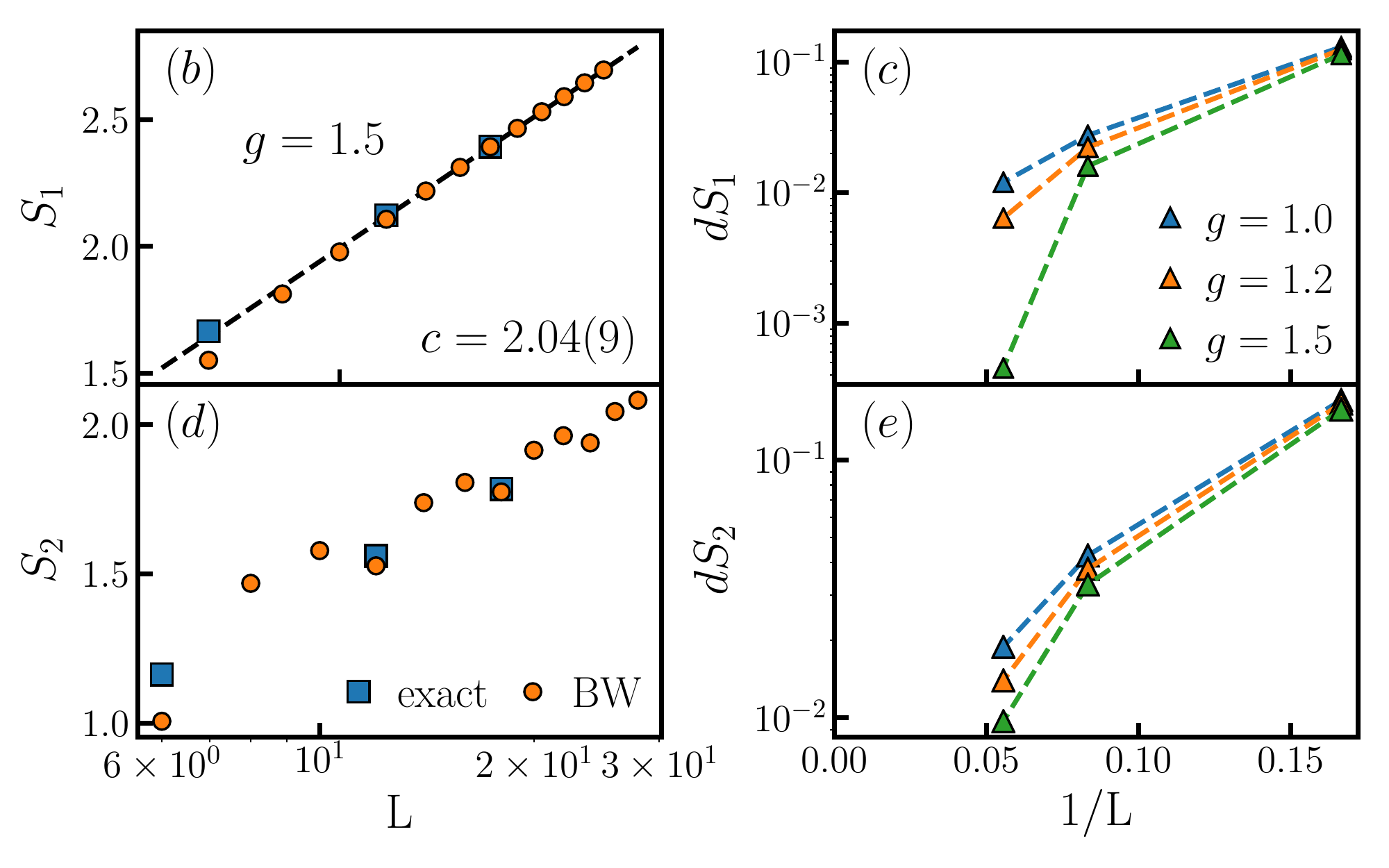}} \ \ \ \ \ \ \
\caption{(a) Sound velocity as defined in Eq.~\eqref{BWvel} for $L_T=6,12,18$. Dashed lines indicate the power-law 
fit with fixed power $2$. Blue horizontal lines are the fitted asymptotic values of $v$, while the green line is 
the exact known value $v = 2 \pi / 3$ for the integrable point $g=1$. The fitted value for these points deviates 
from the exact one with a relative error $< 10^{-6}$. 
Panels (b),(d) show the exact $S_\alpha$ for $\alpha=1,2$  (blue squares) and the BW $S_{\alpha}^{BW}$ (orange circles) for $g=1.5$ (non-integrable point) obtained with ED. The dashed 
line is a fit of the last $5$ points from the BW von Neumann entropy. The resulted central charge is in perfect agreement 
with the expected value, $c = 2$ (c),(e) Discrepancy between $S_\alpha^{BW}$ and the exact $S_\alpha$, as defined in Eq.~\eqref{discr},
for $g=1,1.2,1.5$.} 
\label{fig:bb}
\end{figure}

We employ the BW-EH for finite systems with PBC [Eq.\,\eqref{finitePBC}]. $S_{\alpha}^{BW}$ is obtained for  $L_T \le 22$ using ED.
The sound velocity $v$ can be extracted from the finite size scaling of the ground state energy, assuming the knowledge of the central charge \cite{cardy1986,affleck1986}.
However, here we follow another route. 
Based on the very precise relation between the spectrum of the BW-EH, $\{ \epsilon_{\alpha}^{BW}\}$, and the eigenvalues of $\rho_A$, $\{ \lambda_{\alpha}\}$,
namely\cite{Giuliano2018}
\begin{equation}
 \lambda_\alpha = \frac{e^{-2 \pi \varepsilon^{BW}_\alpha / v  }}{Z_{BW}},
 \label{BWvel2}
\end{equation}
we can compute $v$  from the first two eigenvalues of the BW-EH and $\rho_A$ via the relation
\begin{equation}
v = \frac{ 2 \pi ( \varepsilon^{BW}_1-\varepsilon^{BW}_0 )}{  \log(\lambda_0/\lambda_1) }.
\label{BWvel}
\end{equation}
We verify that Eq.\eqref{BWvel2} holds for $g\ge1$.
First, we compute  $v$ for $g=1$, where the exact sound velocity $v= 2 \pi / 3$ is known from the Bethe ansatz solution. 
Despite we have only 3 points at disposal, 
the extrapolatation of the inifinite-size value of $v$
by fitting $|v(L_T)-2\pi/3|$ with a power law $A/L^\gamma$ ($\gamma = 2$) gives a value for  $v$ that is within 0.5 \% level of the exact result. 
Given the apparent absence of sub-leading corrections, this procedure gives results (see Fig.~\ref{fig:bb} (a)) 
which are more accurate than the values extracted from the finite size scaling of the ground state energy. In fact the latter 
is known to be affected by logarithmic corrections \cite{LUDWIG1987}.
Assuming the aforementioned
power-law scaling, we then extract the value of $v$ for $g>1$ by considering a two-parameter fit on 3 points; see Fig.~\ref{fig:bb} (a)

Given the small system sizes that we can reach with ED, we do not discuss the accuracy of $C_{\alpha}^{BW}$ for the BBM, instead,
we  focus our analyses on the comparison between $S_{\alpha}^{BW}$ and the exact $S_{\alpha}$.
In Fig.~\ref{fig:bb}, we consider $S_\alpha^{BW}$ for $g=1,1.2,1.5$.
As can be seen from Figs.~\ref{fig:bb} (c) and (e), the discrepancy, $dS_{\alpha}$, decreases with system size 
and even for $L = 9$, we observe a discrepancy of just $dS_{\alpha} \approx 10^{-2}$.
This result is observed not just at the integrable point, $g = 1$, but also away from it, i.e., $g = 1.2$ and $1.5$.
Furthermore, the central charge extracted from the size scaling of $S_1^{BW}$ for $g = 1.5$ is in perfect agreement with 
the exact result, $c = 2$, see Fig.~\ref{fig:bb} (b).
Note that there is not sign of frustration felt by exact ground state in the von Neumann entropy computed from the BW-EH.
These results demonstrate that the applicability of the BW-EH ansatz is not restricted to integrable models.

\section{Results II: Bisognano-Wichmann Norm distance}
\label{resultsII}

\begin{figure*}[]
{\centering\resizebox*{8.5cm}{!}{\includegraphics*{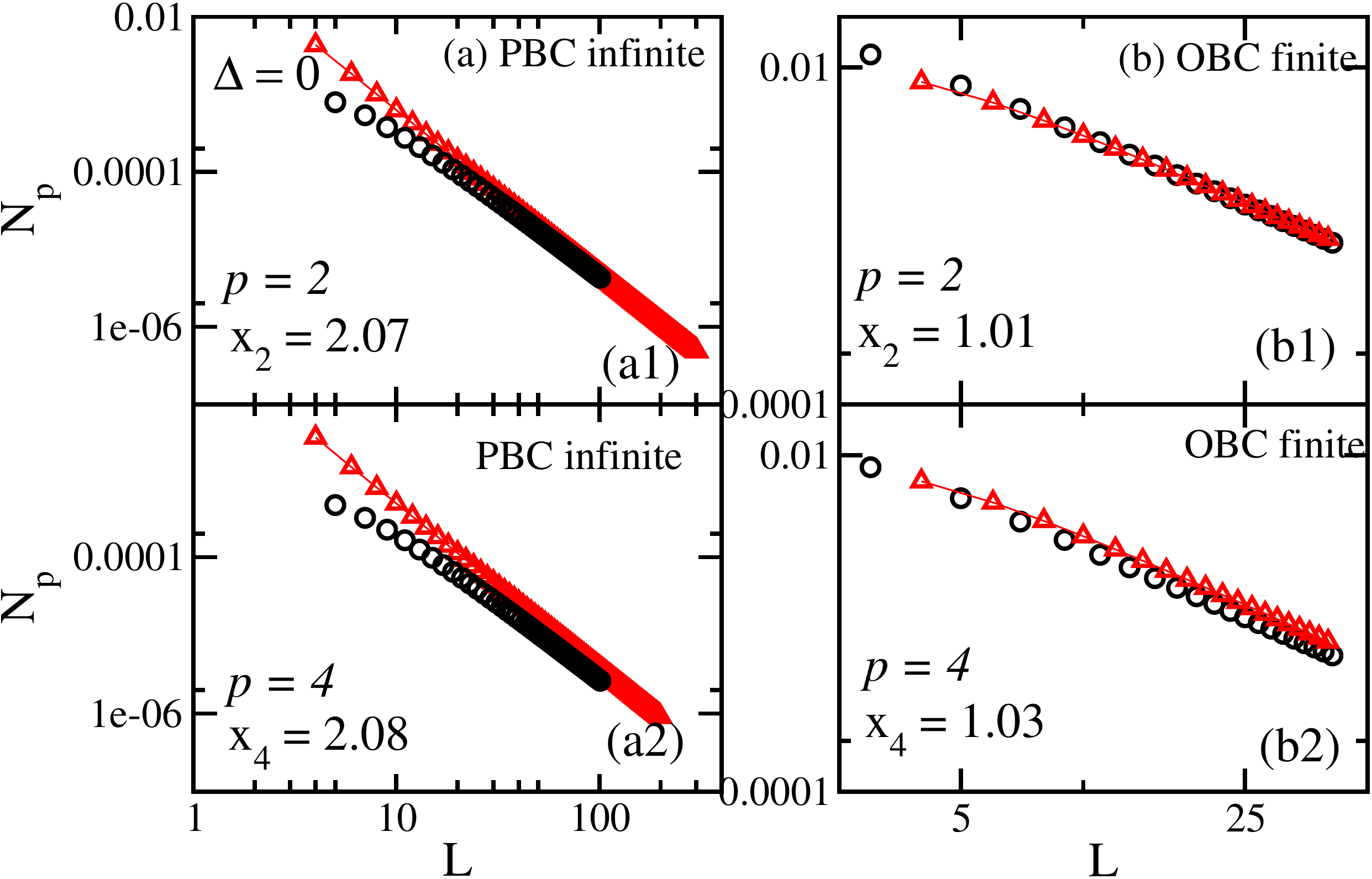}}} \ \ \ \ \ \ \
{\centering\resizebox*{8.6cm}{!}{\includegraphics*{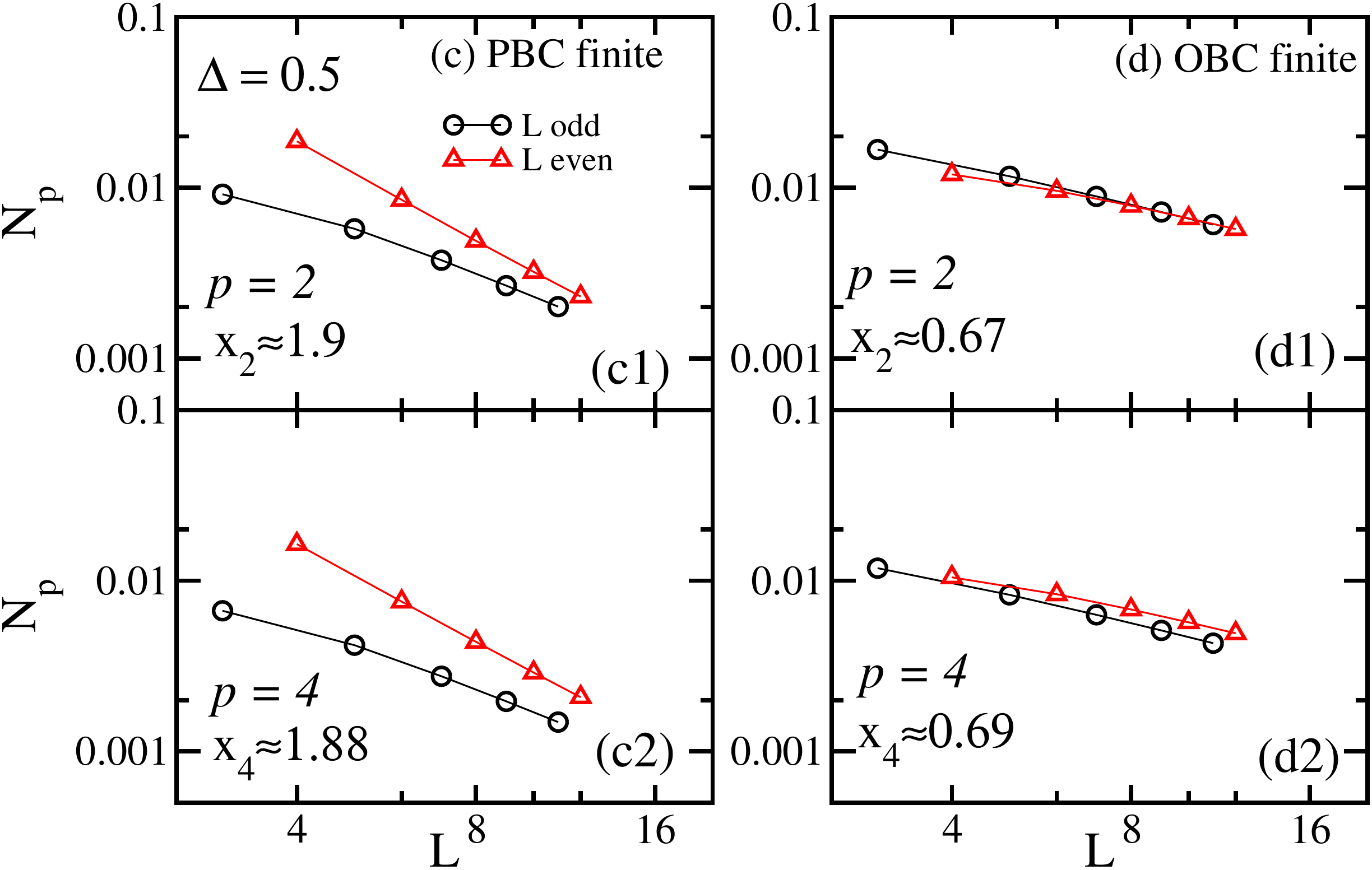}}}
\caption{Schatten norm for the XXZ model with $\Delta = 0$ [panels (a) and (b)] and  $\Delta = 0.5$ [panels (c) and (d)].
Panels (a) and (b) show results for the infinite and finite OBC partitions, respectively, 
obtained by diagonalizing the corresponding free-fermion EH.
Panels (c) and (d) show results for the finite PBC and finite OBC partitions, respectively,
obtained with ED.
The square (black points) and triangles (red points) are results for even and odd values of $L$, respectively.}
\label{fig:norm}
\end{figure*}

\begin{figure}[t]
{\centering\resizebox*{8.6cm}{!}{\includegraphics*{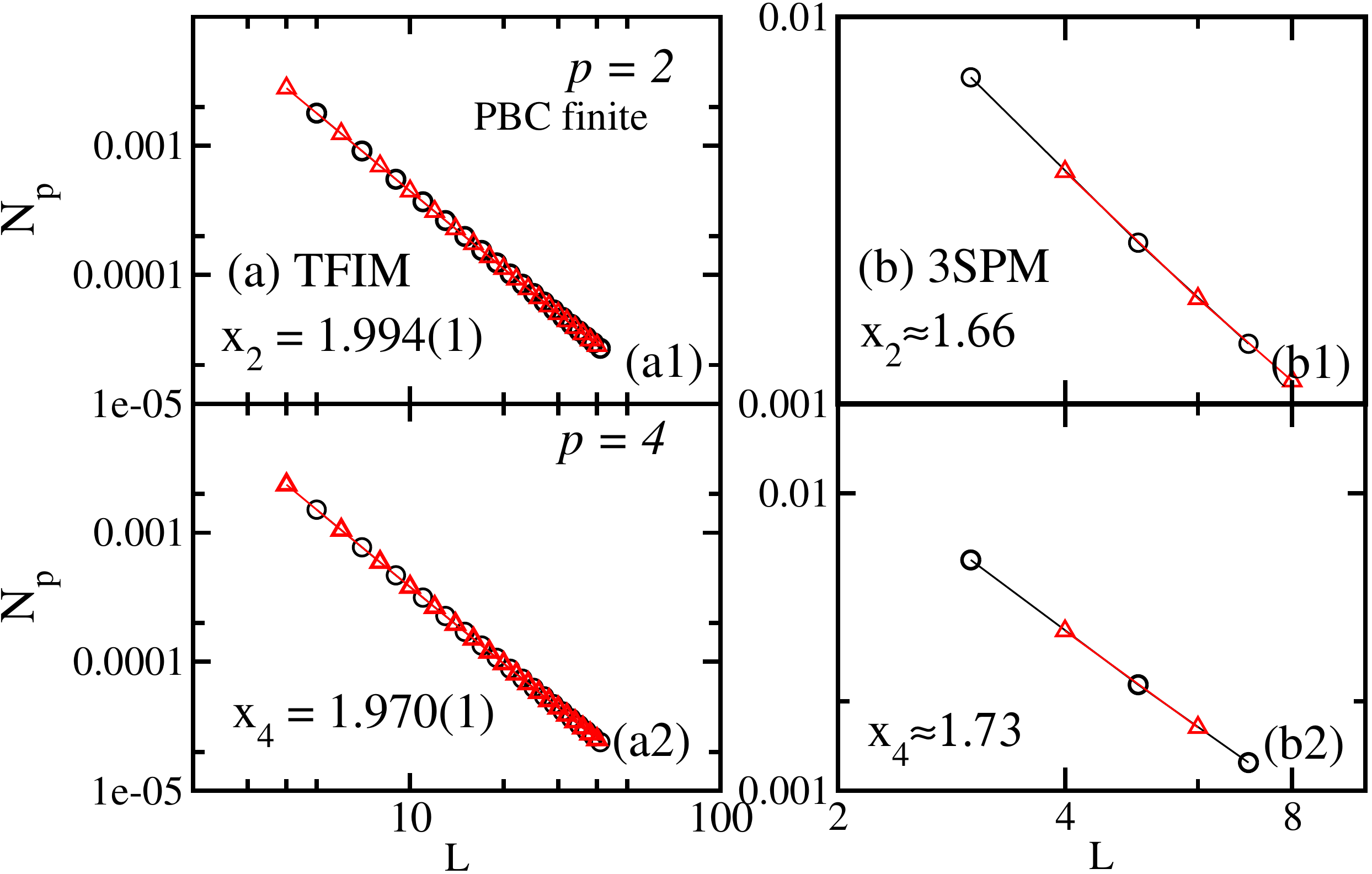}}}
\caption{Schatten norm for the finite PBC partition in the (a) transverse field Ising  and the (b) three-states Potts model.
The square (black points) and triangles (red points) are results for even and odd values of $L$, respectively.
}
\label{fig:normZn} 
\end{figure}

In this section, we discuss the size scaling of the distance  between $\rho_{BW}$ and $\rho_A$, quantified by the Schatten norm $N_p$ [see Eq. \eqref{eq:Schatten}].
Unlike the BW REs, that depends exclusively on the eigenvalues of the BW state,  $\rho_{BW}$,
$N_{p}$ is in general related also to its eigenvectors, and thus probes if indeed $\rho_{BW}$ corresponds to the exact density matrix of the subsystem.
Our discussion is complemented by the analyses of the norm of the  commutator $[\rho_{BW},\rho_{A}]$.
As in the last section, we consider the partitions shown in Fig.\ref{fig:cartoon} (i.e., for finite systems we always consider half-partition).

\subsection{XXZ model}

We start discussing the results for the XXZ model, Eq.\eqref{modelXX}.
For $\Delta = 0$, we use the correlation matrix technique to compute, $\rho_{A}$ \cite{Peschel2004},
and compute $N_p$ for subsystem sizes $L < 10^3$, see Appendix \ref{appendixA}.
For $\Delta \neq 0$ we compute both the exact and the BW reduced density matrices with ED, which allows us to obtain $N_p$  for $L \le 12$. We consider $N_p$ for $p = 2$ (Fobrenius norm) and $p=4$. We remark that, for the $\Delta=0$ case, the Schatten norm is only sensitive to the eigenvalues of the reduced density matrix. 

For $\Delta = 0$,  we consider both  PBC  and OBC.
We observe that  $N_p$ asymptotically goes to zero as a power law
\begin{align}
 N_{p} \sim \frac{1}{L^{x_p}},
 \label{power}
\end{align}
which confirms that $\rho_{BW}$ is exact in the thermodynamic limit, see Fig. \ref{fig:norm} (left graph).
Moreover, it is worth to point out two interesting features of this power-law decay.
First,  $N_p$ exhibits an oscillating behavior with $L$ for the PBC case. 
Second,  by fitting the results  of Figs. \ref{fig:norm} (a) and (b) (we just consider even values of $L$) with $P(L) = A/L^{x_p}$,
we observe that the scaling exponent, $x_p$, is independent of $p$, and is given
by $x_{2} \approx 2$ ($x_{4} \approx 2$)  and $x_{2}  \approx  1$ ($x_{4} \approx 1$) for the  PBC and OBC, respectively.
Remarkably,  these values  indicates that $x_p$ is compatible with  $x_p = \eta K$, where $\eta =1,2$ for OBC/PBC,
and $K$ is the Luttinger liquid parameter;
which suggests that $x_p$ is related to the scaling dimension of the energy density operator,
as it occurs with the corrections to the CFT R\'enyi entropy, discussed in the last section. 

For the interacting case ($\Delta = 0.5$), $N_p$ decays as $L$ increases, and  even for subsystem sizes $L = 10$, $N_p \approx 10^{-3}$,  see Fig. \ref{fig:norm} [panels (c)-(d)].
This result strongly indicates that $N_p \to 0$ in the limit of $L\to\infty$.
Curiously, $N_p$ exhibits parity oscillations with $L$ in the PBC case, and
the $x_p$ obtained by fitting $N_p$ (we just consider even values of $L$) with $P(L) = A/L^{x_p}$ is consistent with 
a $p$-independent $x_p$.
In this case, despite our results indicate that the value of $x_p$ depends on the BC, 
as suggested by the results of the XX model, we are not able to confirm if indeed 
$x_p=\eta K(\Delta)$, due to the limitation of system sizes that we can reach with ED.
As shown in Fig. \ref{fig:norm} [panels (c)-(d)] , the values of $x_p$ obtained for OBC have a discrepancy of approximately $8\%$ 
with respect to $x_p =  K(0.5)$. For PBC this discrepancy is of order $20\%$, when we just consider even values of $L$ in the fit.

\begin{figure*}[]
{\centering\resizebox*{8.45cm}{!}{\includegraphics*{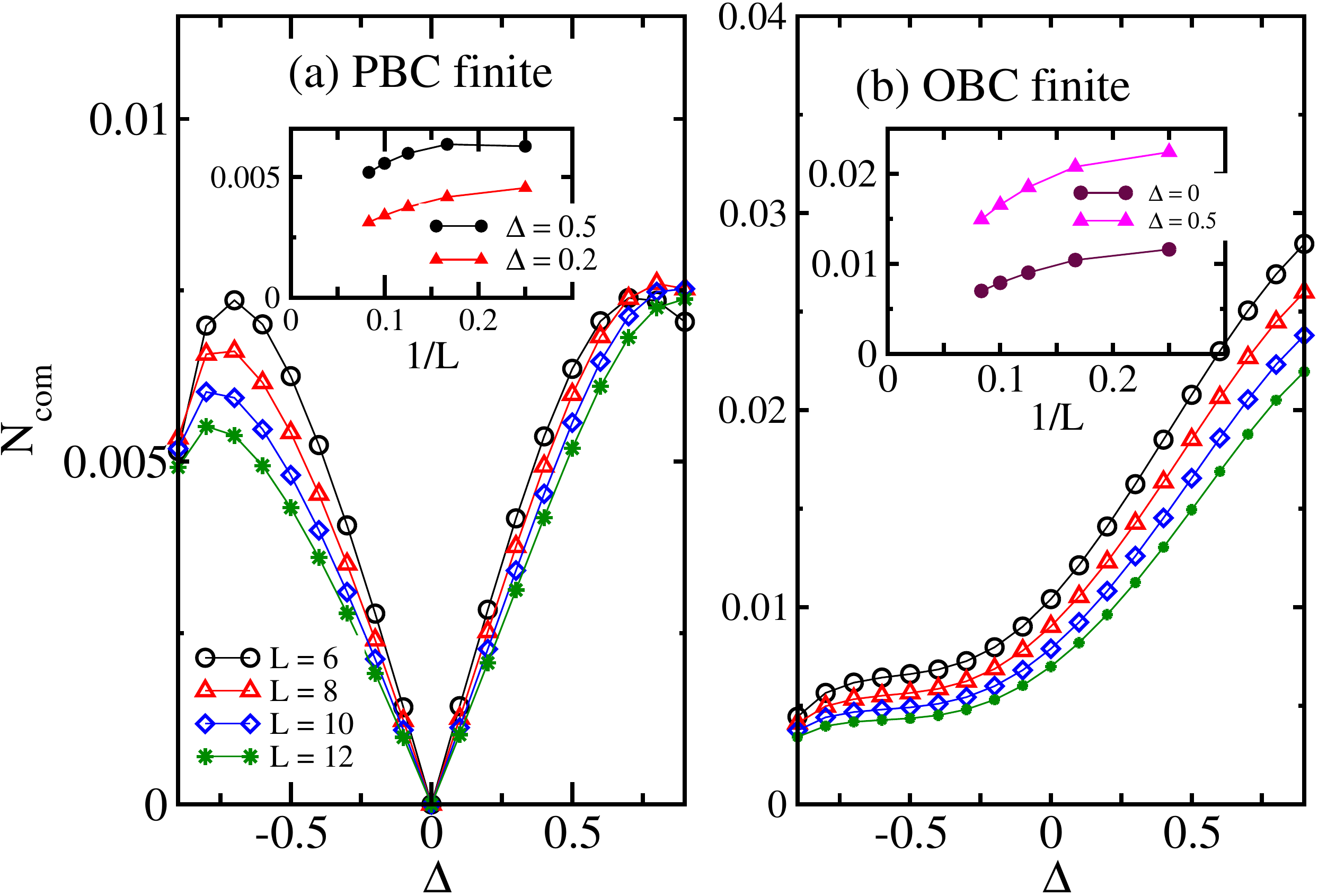}}} \ \ \ \ \ \ \
{\centering\resizebox*{8.6cm}{!}{\includegraphics*{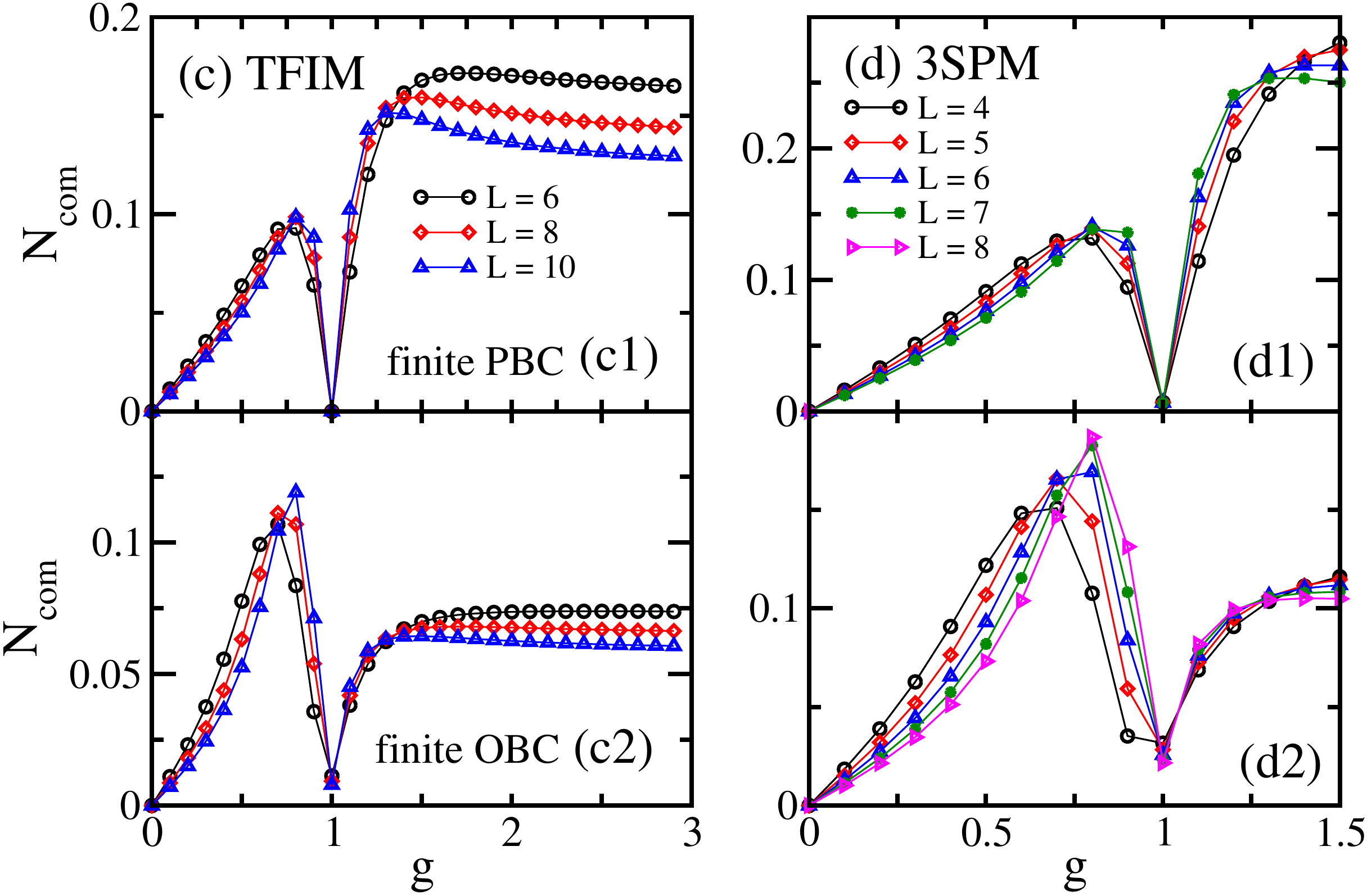}}}
\caption{\textit{Commutator between the exact reduced density matrix and the BW-EH}. Panels (a) and (b): results  for the XXZ model.  Panels (c) and (d): results for  the transverse field Ising and three-states Potts models,
respectively.
The results are obtained with ED.}
\label{fig:commu}
\end{figure*}

\subsection{Transverse field Ising and quantum three-states Potts models}

We now consider $N_{p}$ for the discrete-symmetric models, i.e. TFIM and 3SPM.
In these cases we focus on the finite PBC partition,
the results for the OBC case are similar.
For the TFIM, we use the correlation matrix technique to compute $\rho_{A}$ (see Appendix \ref{appendixA})
which allows us to compute $N_p$ for subsystem sizes $L \sim 10^2$.
For the 3SPM case, we compute both the exact and the BW reduced density matrices with ED, which allows us to obtain $N_p$  for $L \le 8$.

As it occurs for the XXZ model, our results strongly indicate that for both models $\rho_{BW}$ is precise in the thermodynamic limit; see Fig. \ref{fig:normZn}.
Unlike what is observed for the XXZ model, and similarly to the size scaling of the  REs, $N_{p}$ does not exhibit any oscillatory behavior with respect to $L$.
However, the exponents $x_p$ are close to $x_p = \eta X_e$.
The values obtained for $x_p$ is within $1\%$ ($6\%$) level with the scaling exponent of the energy density operator for the TFIM (3SPM) [$X_e = 1$ for the TFIM ($X_e = 4/5$ for the 3SPM)]; see Fig. \ref{fig:normZn}.
These results suggest that $x_{p}$ may be related to the scaling dimension of the energy density operator.

Before closing this subsection, it is important to mention  that a possible problem in using $N_p$ [Eq. \eqref{eq:Schatten}] as measure of
distance between two reduced density matrices	 is that, in general, even the Schatten norm of the operators $\rho_A$ and $\rho_{BW}$
goes to zero when $L \to \infty$.
Thus, the fact that $N_2~\to~0$ should not be conclusive about the distance between $\rho_{A}$ and $\rho_{BW}$ in the asymptotic limit, i.e.,
in principle, $N_2 \to 0$ simply because the  the Schatten norm of  $\rho_A$ (and $\rho_{BW}$) itself goes to zero.	
Nevertheless, the main conclusion drawn above  is not an artefact of this issue, as we explain bellow.
For $N_2$, for instance, one can consider the normalized norm,
$\tilde{N}_2 = N_2/ K_2$\cite{Fagotti2013}, where
\begin{equation}
K_2 = \sqrt{\tr{\rho_{A}^2} + \tr{\rho_{BW}^2}} =  \sqrt{e^{-S_2} + e^{-S_2^{BW}}},
\end{equation}
in order to circumvent the aforementioned problem.
The asymptotic behavior of this normalization factor can be related to the central charge;
considering the expression for the Renyi entropy [Eq. \eqref{ScftFinite}], we obtain $K_2 \sim 1/L^{c/(8\eta)}$, where $\eta = 1,2$ for PBC/OBC.
Thus, the asymptotic behavior of $\tilde{N}_2 $ is given by
\begin{equation}
 \tilde{N}_2 \sim \frac{1}{L^{x_2 - c/8\eta }}.
\end{equation}
As $x_2 > c/(8\eta)$ for all the cases discussed above, we also observe that $\tilde{N}_2$ asymptotically goes to zero as a power law.

\subsection{Norm of the BW commutator}

In the context of free-fermion models (i.e., for the XX model and the TFIM) it is known that 
the BW-EH for  the finite or infinite PBC partitions, see Fig.\ref{fig:cartoon} (a) and (b), commutes with the exact EH and $[\rho_{BW},\rho_{A}] = 0$.
In this context, the BW-EH is usually referred to as $T$ matrix, see Refs.~\onlinecite{Peschel2004,Peschel2018}.
An important consequence of this result is that the norm distance between  $\rho_{BW}$ and $\rho_{A}$ depends exclusively on their eigenvalues (not on their eigenvectors), as can be deduced from the definition of $N_p$.
Hence, the fact that $N_{p}$ asymptotically goes to zero as a power law, and even more intriguing, that this power law seems related to the scaling dimension of the energy density operator, is solely due to the properties of the eigenvalues of $\rho_{BW}$ and $\rho_{A}$.

For the finite OBC case, the $T$ matrix is not exactly equal to the BW-EH, despite being very close to it.
In fact, in this case, the $T$ matrix is obtained by a simple change, $2L\to 2L+1$, in Eq. \eqref{finiteOBC} \cite{Peschel2018}.
However, for the interacting models considered here, i.e. XXZ ($\Delta \neq 0$) and 3SPM, it is not known whether a discrete $\rho_{BW}$ such that $[\rho_{BW},\rho_{A}] = 0$ exists.
In order to investigate these issues, we now discuss the behavior of the Fobrenius norm of the commutator 
\begin{equation}
 N_{com} = \sqrt{\tr {([\rho_{A},H_{BW}][\rho_{A},H_{BW}]^{\dagger})}}.
\end{equation}
We consider the commutator $[\rho_{A},H_{BW}]$, instead of $[\rho_{A},\rho_{BW}]$, to single out the effects of 
$\beta_{EH}$, especially when the system is away from the critical point (see bellow).

\subsubsection{XXZ model}

We first consider $N_{com}$ as a function of $\Delta$ for the XXZ model [see Eq. \eqref{modelXX}].
For the finite PBC partition, $N_{com}$ is numerically equal to zero  at $\Delta = 0$, which is consistent with the results of the $T$ matrix \cite{Peschel2018};
see Fig.\ref{fig:commu} (a) (left graph).
For $\Delta \neq 0$,  $N_{com} \neq 0$, however, $N_{com}$ decreases with $L$, as shown in the inset of Fig.\ref{fig:commu} (a).

For the finite OBC partition, even at the free-fermion-model point ($\Delta = 0$) $N_{com} \neq 0$,
which is in agreement with the fact that the BW-EH does not correspond to the $T$ matrix in this case.
Nevertheless, $N_{com}$ decreases with the subsystem size in all the parameter region, $ -1 < \Delta \le 1$,
suggesting that $N_{com} \to 0$  as $L \to \infty$ in the whole gapless critical region
of the XXZ model.

\subsubsection{TFIM and 3SPM}

We now consider the behavior of $N_{com}$ for the TFIM and 3SPM as function of the coupling $g$, see Eq.\eqref{modelIsing} and \eqref{H3pm}.
For the TFIM,  we observe a sharp dip of $N_{com}$ at the quantum critical point, $g_c=1$, see Fig.\ref{fig:commu} (c).
In fact, for the finite PBC, $N_{com}$ is numerically equal to zero at $g = g_c$, as expected for the corresponding $T$ matrix \cite{Peschel2018}.
For the finite OBC, despite $N_{com}$ not being exactly equal to zero at $g_c$, its value is negligible compared to
the results for $g \neq g_c$ (e.g., for $L = 10$, $N_p \approx 10^{-3}$ at $g_c$). 
Furthermore, it decreases with the subsystem size at $g_c$; see Fig.\ref{fig:commu} (c2),
which indicate that $N_{com} \to 0$ in the thermodynamic limit, as observed for the XXZ model. 

Remarkably, we observe a similar result for $N_{com}$ at the critical point of the interacting 3SPM, as can be seen in Fig.~\ref{fig:commu} (d).
For the finite PBC partition, we can barely see from Fig.\ref{fig:commu} (d1), that $N_{com}$ is not exactly equal to zero at $g_c$  ($N_{com} \approx 10^{-3}$ in this case).

Summing up our results, we observe that for all the critical models considered here, $N_{com}$ is
almost negligible when compared to the $N_{com}$ of non-critical models (i.e., $g \neq g_c$), moreover $N_{com} \to 0$ as $L \to \infty$.
These results suggest that $[\rho_{BW},\rho_{A}] \approx 0$ might be a general feature  of  quantum critical chains.
Consequently, in these cases the norm distance $N_{p}$ depends exclusively on the eigenvalue properties of both $\rho_{BW}$ and $\rho_{A}$.

\section{Discussion and conclusions}
\label{conclusion}

In this paper we presented an extensive numerical investigation of the accuracy of the BW R\'enyi entropy, $S_{\alpha}^{BW}$,
and the BW state itself (i.e., $\rho_{BW}$) for one-dimensional critical models.
Our results are based on different numerical techniques, including exact diagonalization, DMRG and QMC.
Our conclusions are: for all the models considered here, we observe that 
(i) both $S_{\alpha}^{BW}$ and  $\rho_{BW}$ converge to the  exact results in the thermodynamic limit,
and,  $S_{\alpha}^{BW}$ describes (ii) not just the CFT logarithmically-divergent term, but also some universal lattice-finite-size  corrections to the CFT formula.
For the last point, we show that the power-law decay of the leading term of $C_{\alpha}^{BW}$ is related to the scaling dimension of the energy density operator $p_{\alpha}$ (with the exception of $\alpha = 1$ for PBC).

In Ref.~\onlinecite{Tonni2019} it was recently shown that the exact lattice EH of free-fermion chains at half-filling (XX model) 
is equal to the conformal expression [Eq.~\eqref{BWtheorem} with the appropriate $\lambda(n)$] if one
includes the hopping at finite distance in the continuum limit of the entanglement lattice Hamiltonian.
This result shows that the tiny long-range terms present in the exact EH, but absent in the BW-EH, are irrelevant terms in the asymptotic limit,
and explains why the BW REs are remarkably close to the exact results in the thermodynamic limit.
Our observation that $dS_{\alpha} \to 0$ for the XXZ and 3SPM models indicates that \textit{possible} long-range terms presented in the exact EH
of these critical interacting lattice models are also irrelevant in the thermodynamic limit.

On the other hand, the observation that $S_{\alpha}^{BW}$ properly describes universal lattice-finite-size effects
associated to the scaling exponent $p_{\alpha} = \eta X_{e}/\alpha$, can be understood if one considers the 
conformal mapping used to obtain the EH of the partitions considered here \cite{Cardy2016}.
Remarkably, we observe that even the coefficient of these corrections are almost equal to the exact ones.
We thus conclude that \textit{the almost negligible} discrepancy $dS_{\alpha}$ is related to subleading 
corrections presented in both the exact and the BW $C_{\alpha}$.
From a methodological point of view, this demonstrate that our approach may be used to check convergence of tensor network states in conformal phases, especially for large values of the central charge (since the complexity of the Wang-Landau method is not affected by the entanglement of the ground state wave function).

The discussion we have presented here only concerns
one-dimensional critical lattice models  whose Hilbert space can be written in tensor product form.
A possible extension of our approach includes 1D critical points described by models whose Hilbert space is subjected to constraints \cite{Sachdev2004,Chepiga:2019aa,Samajdar:2018aa,Angelone2019,Morampudi:aa}.
Another important feature of our approach, is that it  is restricted to ground state properties.
It would be interesting to extend it to thermal EH \cite{Assad2018,Cardy2016},
which could shed light on the open problem of measuring entanglement on thermal states.

\acknowledgements{We acknowledge useful discussions with P. Calabrese, B. Vermersch and V. Eisler.
This work is partly supported by the ERC under grant number 758329 (AGEnTh), and has received funding from the European Union's Horizon 2020 research and innovation programme under grant agreement No 817482. 
TMS and MD acknowledge computing resources at Cineca Supercomputing Centre through the Italian SuperComputing Resource Allocation via the ISCRA grants QMCofEH.
MAR thanks ICTP for the financial support and hospitality. The work of MAR is partially supported by CNPq.}


\phantomsection
\addcontentsline{toc}{chapter}{Bibliography} 
\bibliography{EntCorrection_v2}

\appendix

\section{Renyi entropy and distance norm of free-fermion entanglement Hamiltonians}
\label{appendixA}

In this Appendix we discuss further details of the calculation of the BW REs and the distance norm for free-fermion models.

Using the Jordan Wigner transformation we can map both the XX [Eq.\eqref{modelXX} with $\Delta = 0$] and the transverse field Ising [Eq.\eqref{modelIsing}] models in spinless free-fermion Hamiltonians.
The corresponing BW-EH are then given by
\begin{equation}
 H_{BW}^{XX} = -\frac{J}{2} \sum_{n=1}^{L-1} \lambda(n) \left( c^{\dagger}_n c_{n+1} + h.c. \right),
 \label{nonIntXX}
\end{equation}
and
\begin{align}
 H_{BW}^{TFIM} =&-\sum_{n=1}^{L-1} \lambda(n)  \left( c^{\dagger}_n c_{n+1} + c^{\dagger}_{n+1} c_{n}   + c^{\dagger}_n c_{n+1}^{\dagger} + c_{n+1} c_{n}\right) \nonumber \\
	       &-2g \sum_{n=1}^{L}  \lambda(n-1/2)  c^{\dagger}_n c_{n}  + K
 \label{nonIntTIM}
\end{align}
for the XX model and the TFIM, respectively.
The $c_{n}$`s and $c_{n}^{\dagger}$`s are Fermi annihilation and creation operators, $K$ is a constant,
and the couplings $\lambda(n)$ [see Eqs.~\eqref{infiniteOBC},\eqref{infinitePBC},\eqref{finiteOBC},\eqref{finitePBC} of the main text] define the geometry  of the partitions, see Fig. \ref{fig:cartoon}.
When $\lambda(n)$ is given by the Eqs.\eqref{infinitePBC} and \eqref{finitePBC} the BW-EH commute with the exact EH \cite{peschel2009,Peschel2018}.
By diagonalizing these quadratic BW-EH through  a canonical transformation one can obtain 
the corresponding spectrum. 
The BW REs [see Eq. \eqref{BWrenyi}] are then calculated by using the definition of the thermodynamic free energy for non-interacting fermions [see Eq. \eqref{eq:freeQ}].

The Schatten norm is defined as
\begin{align}
N_p = \left(\tr D^p\right)^{1/p}, 
\label{eq:norm}
\end{align}
where the distance, $D$, is
\begin{equation}
D = \rho_{BW} - \rho_{A}.
\label{eq:D}
\end{equation}
The exact reduced density matrix
\begin{equation}
 \rho_{A} = \frac{e^{-H_{A}}}{Z_A},
\end{equation}
and the exact EH, $H_A$, are calculated using the correlation matrix technique developed in the Ref.\cite{Peschel2004}.
It is important to mention that $H_{A}$ is also a quadratic fermionic Hamiltonian.

The Schatten norm $N_p$ is obtained with the aid of the following relation
\begin{align}
 \tr\left(e^{-\hat{H}}\right) = \det\left(1 + e^{-H}\right),
\end{align}
where   $\hat{H} = \vec{c} \ H  \ \vec{c}^{T}$ is quadratic fermionic Hamiltonian, 
$\vec{c} = (\hat{c}_1^{\dagger},\hat{c}_2^{\dagger}, ..., \hat{c}_L^{\dagger})$, and $H$ is a $L \times L$ matrix.
For example, for $p=2$, we  can write
\begin{align}
  \tr D^{2} =  & \frac{\det\left( 1 + e^{-2 \tilde{H}_{BW}} \right)}{\left[\det\left( 1 + e^{-\tilde{H}_{BW}} \right) \right]^2} 
  + \frac{\det\left( 1 + e^{-2 H_{A}} \right)}{\left[\det\left( 1 + e^{-H_{A}} \right) \right]^2} + \nonumber \\
  & \frac{\det\left( 1 + e^{-H_{A}} e^{-\tilde{H}_{BW}} \right)}{\det\left( 1 + e^{-H_{A}} \right) \det\left( 1 +  e^{-\tilde{H}_{BW}} \right)},
  \label{eq:D}    
\end{align}
where $\tilde{H}_{BW} = \beta_{EH} H_{BW}$. 
Similar expressions can be obtained for  $p=4$.

\end{document}